\documentclass[aps,jmp,amsmath,amssymb,preprint]{revtex4-1}
\usepackage{dcolumn}
\usepackage{bm}
\usepackage{graphicx}
\usepackage{subfigure}
\usepackage{physics}
\usepackage{subscript}
\usepackage{soul}
\usepackage[utf8x]{inputenc} 
\usepackage{verbatim}
\usepackage[usenames,dvipsnames]{color}

\begin{document}

\title{Comprehensive studies on steady-state and transient electronic transport in  In\textsubscript{0.52}Al\textsubscript{0.48}As}

\author{Anup Kumar Mandia$^{1}$, Bhaskaran Muralidharan$^{1}$},
\author{Seung Cheol Lee$^2$}
\email{seungcheol.lee@ikst.res.in}
\author{Satadeep Bhattacharjee$^{2}$}
\email{satadeep.bhattacharjee@ikst.res.in}
\affiliation{$^{1}$Department of Electrical Engineering, Indian Institute of Technology Bombay, Powai, Mumbai-400076,  India \\
$^{2}$Indo Korea Science and Technology Center, Bangalore-560065, India}

\begin{abstract}
High electron mobility transistors (HEMT) built using In\textsubscript{0.52}Al\textsubscript{0.48}As/In\textsubscript{0.53}Ga\textsubscript{0.47}As on InP substrates are a focus of considerable experimental studies due to their favourable performance for microwave, optical and digital applications .
We present a detailed and comprehensive study of steady state and transient electronic transport in In\textsubscript{0.52}Al\textsubscript{0.48}As with the three valley model using the semi-classical ensemble Monte Carlo method and including all important scattering mechanisms. All electronic transport parameters such drift velocity, valley occupation, average electron energy, ionization coefficient and generation rate, electron effective mass, diffusion coefficient, energy and momentum relaxation time are extracted rigorously from the simulations. Using these, we present a complete characterization of the transient electronic transport showing the variation of drift velocity with distance and time. We have then estimated the optimal cut-off frequencies for various device lengths via the velocity overshoot effect. Our analysis shows that for device lengths shorter than $700$ nm, transient effects are significant and should be taken into account for optimal device designs. As a critical example,  at length scales of around $100$ nm, we obtain a significant improvement in the cut-off frequency from $261$ GHz to $663$ GHz with the inclusion of transient effects. The field dependence of all extracted parameters here can prove to be helpful for further device analysis and design.       
\end{abstract}


\maketitle
\section{Introduction}
High electron mobility transistors (HEMT) built using heterostructures of
In\textsubscript{0.52}Al\textsubscript{0.48}As/In\textsubscript{0.53}Ga\textsubscript{0.47}As 
on InP substrates are a focus of a great deal of experimental 
studies due to their favourable performance in microwave, optical and digital 
applications 
\cite{app1,app2,app3,app4,app5,app6,app7,app8,app10,app11,app12,app13,app14,app15,Rev}. 
Laser\cite{app16} and charge injection\cite{app17} transistors (CHINT) fabricated from 
such material systems have also shown promising device performance characteristics such 
that InAlAs/InGaAs/InP structures are considered to be among the best owing to better 
integration, coupled with higher power efficiency, faster speed, high frequency gain, 
lower noise all coupled with low cost. \\
\indent HEMTs based on In\textsubscript{0.52}Al\textsubscript{0.48}As/In\textsubscript{0.53}Ga\textsubscript{0.47}As/InP have a cut-off frequency higher than 600 GHz and are considered to be among the fastest transistors \cite{fast1,fast3}. These HEMTs have also shown significant capabilities for cryogenic operations in terms of improved noise characteristics \cite{Rev,cry1,cry2,cry3,cry4,cry5,cry6,cry7,cry8}. Such HEMT structures coupled with GaAs and InAs have also shown good device performances \cite{per1,per2,per3,per4,per5,per6,per7,per8} suitable for various digital and analog applications. To exploit the application potential from these structures, a deeper understanding of transport across such structures is required. There is a lot of experimental and theoretical studies done for InP and In\textsubscript{0.53}Ga\textsubscript{0.47}As, but almost very little information is available for the In\textsubscript{0.52}Al\textsubscript{0.48}As system in terms of material parameters and transport properties. The objective of this paper is to present a detailed and comprehensive study of steady state and transient electronic transport in In\textsubscript{0.52}Al\textsubscript{0.48}As with the three valley model using the semi-classical ensemble Monte Carlo method and including all important scattering mechanisms. In our model, all electronic transport parameters such drift velocity, valley occupation, average electron energy, ionization coefficient and generation rate, electron effective mass, diffusion coefficient, energy and momentum relaxation time are extracted rigorously from the simulations. \\
\indent There are a wide variety of semi-classical transport models \cite{drift,hydro1,hydro2, drift_maxw, sp_harm1, sp_harm2, sp_harm3, sp_harm4, sp_harm5, sp_harm6, sp_harm7, Rode, Rode1, Anup1, variational1, variational2, alternative1, alternative2, Monte1, Monte2, Monte3, Monte4, Monte5, Monte6, automata1, automata2, weighted1, weighted2, matrix1, matrix2, matrix3, iterative1, iterative2, iterative3} being used to understand transport physics. Of them all, the Monte Carlo method \cite{Monte1,Monte2} is considered to be the most accurate which is much easier to implement and provides better insights from a physical point of view. For many years a lot of papers have been published based on Monte Carlo technique \cite{Monte1, Monte2, Monte3, Monte4, Monte5, Monte6} for transport properties calculations of different materials. The accuracy of these methods are limited only by the models used to calculate the band structure and the scattering rate. \\
\indent For larger device dimensions, an understanding of steady-state transport is sufficient. In smaller devices, transient transport is also essential when estimating device performance. Transient electronic transport in short channel FETs was first studied by Ruch \cite{Ruch}. It was shown that transient electron drift velocity may exceed steady state drift velocity by proper selection of electric field. Heiblum \cite{Heiblum} had done first experimental observation of transient electron transport for GaAs. Later a lot of investigations were done both theoretically and experimentally for transient transport \cite{trans1,trans2,trans3,trans4,trans5} for different materials.\\
 \indent Electronic transport properties in In\textsubscript{0.52}Al\textsubscript{0.48}As earlier analysed only in the steady state \cite{earlier1,earlier2} by using Monte Carlo methods. A previous study has also shown that electron transit time in In\textsubscript{0.52}Al\textsubscript{0.48}As layer is the main factor that decides the total device transit time in In\textsubscript{0.52}Al\textsubscript{0.48}As based CHINT\cite{app16}. So for the fabrication of high-speed devices, ultra short devices, the analysis of transient electronic transport is necessary. \\     	
\indent In this work, initially we focus on the examination of steady state transport, by studying the variation of drift velocity with electric field, temperature and doping concentrations. Next the variation in electron energy, electron occupancy in different bands with electric field is discussed. The variation of impact ionization coefficient and the generation rate with electric fields also examined further. The diffusion coefficient, momentum and energy relaxation variation with electric field and temperature are presented. Using these, we present a complete characterization of the transient electronic transport showing the variation of drift velocity with distance and time. We have then estimated the optimal cut-off frequencies for various device lengths via the velocity overshoot effect. Our analysis shows that for device lengths shorter than $700$ nm, transient effects are significant and should be taken into account for optimal device designs. As a critical example,  at length scales of around $100$ nm, we obtain a significant improvement in the cut-off frequency from $261$ GHz to $663$ GHz with the inclusion of transient effects. At last device implications of our results are discussed and the upper bound cut-off frequencies for device optimization are calculated for short channel high frequency electronic devices.   \\
\indent This paper is organized as follows: In the following section, the Monte Carlo procedure is discussed and the related parameters required to study transport in In\textsubscript{0.52}Al\textsubscript{0.48}As are presented in detail. Furthermore, the method for calculating the diffusion constant, momentum and energy relaxation times are discussed. In Sec. III, the results of our simulations are discussed thoroughly. First, we discuss the velocity-field characteristics in n-type In\textsubscript{0.52}Al\textsubscript{0.48}As for different temperatures and doping concentrations. Then, the variation of diffusion coefficient, momentum and energy relaxation times with electric field and temperature are presented. Next, the transient electronic transport that occurs in In\textsubscript{0.52}Al\textsubscript{0.48}As with both distance and time are discussed. At last, a device implication of our results are commented upon. Finally, Sec. IV summarizes the important conclusions of this paper. 

\section{Simulation Setup and Formulation}
\subsection{Monte Carlo Procedure}
\indent We have studied the electron transport in bulk In\textsubscript{0.52}Al\textsubscript{0.48}As using ensemble Monte Carlo method. We have used a three-valley model for the conduction band structure of the electrons.
For Monte Carlo simulation we have used a time step of ten femtoseconds and for steady state analysis we have done simulation for 100 picoseconds. Further details of Monte Carlo method is given in references \cite{Monte2, Monte4, Jacoboni-1-review}. Band structure is treated by using non parabolic band structure \cite{conwell}. The dispersion relationship is given by

\begin{equation}
E(k)(1+\alpha E(k)) = \gamma(E(k)) =\frac{\hbar^2 k^2}{2m^*}
\label{band_struct}
\end{equation}

where $ k $ is the wave vector, $ E(k) $ is the energy of a particle of wave vector $ k $, $ \hbar $ is the reduced Planck 
constant, $ \alpha $ is non parabolic coefficient and it is given by

\begin{equation}
\alpha = \frac{1}{E_g}(1-\frac{m^*}{m_0})^2
\label{non_para}
\end{equation}

where $E_g$ is energy band gap, $m^*$ is effective mass of electron at the bottom of the band and $m_0$ is free electron mass.  

For both steady state and transient analysis ten thousand electrons are considered. We assume that all donors are ionized and free electron concentration is equal to the donor concentration. In all cases the doping concentration is set to $1 \times 10^{22} m^{-3}$ for our simulation unless doping concentration is mentioned explicitly. 
     
The material parameters used in the calculation for bulk In\textsubscript{0.52}Al\textsubscript{0.48}As are listed in the table \ref{table1} and \ref{table2}. For required alloy composition, all values are linearly extrapolated between the material parameters of AlAs and InAs \cite{databook}.        

\begin{table}
\caption{ The material parameters for bulk In\textsubscript{0.52}Al\textsubscript{0.48}As}
\label{table1}
\begin{tabular}{lccccccc}
\multicolumn{2}{c}{}   \\
Parameter  &  Value       \\
\hline
Bulk Material Parameters  \\                                                  
Polar Optical Phonon Energy $(eV)$     &   $ 0.0397 $  \\
Low frequency dielectric Constant $ \epsilon_{s} $     &   $ 12.414 $   \\
High Frequency Dielectric Constant $ \epsilon_{\infty} $     &   $ 10.072 $  \\
Energy Band Gap $ E_g (eV) $        &   $ 1.44 $   \\
Density $ \rho $ $(kg/m^3)$              &   $ 4753 $     \\
Acoustic Deformation Potential $ D_{ac} (eV) $          &   $ 7.936 $  \\
Sound Velocity $ v_s (m/s)$    &   $ 4.998 \times 10^3 $   &      \\
Piezoelectric constant $(P_{pz})$    &   $ 0.048069 $   &        \\
Alloy Scattering Potential $(eV)$      &   $ 0.47 $   &        \\
\hline
Elastic constants \\
$ c_{11} $ $(N/m^2)$   &   $ 1.01 \times 10^{11} $     \\
$ c_{12} $ $(N/m^2)$    &   $ 5.11 \times 10^{10} $     \\
$ c_{44} $ $(N/m^2)$      &   $ 4.78 \times 10^{10} $     \\
\hline
\end{tabular}
\end{table}

\begin{table}
\caption{Valley Dependent Parameters for bulk In\textsubscript{0.52}Al\textsubscript{0.48}As}
\label{table2}
\begin{tabular}{lccccccc}
\multicolumn{4}{c}{}   \\
Parameters &  $ \Gamma $  &  L  &  X    \\
\hline
Effective Mass  $ m^{*} $ &   0.08396    &   0.39   & 0.602     \\

Non-parabolicity $ \alpha (eV^{-1}) $ &  0.58273  &  0.20904  &  0.066556             \\

Valley Separation $ (eV) $  & ---  & 0.34  &  0.6     \\

Number of Equivalent valleys  & 1  &  4   &  3         \\

Optical phonon Energy \\
$E_{op} (eV)$   &  $ 0.0397 $  &  $ 0.0397 $  &  $ 0.0397 $        \\

Intervalley Deformation \\
potential $ D_i (eV/m) $  \\
From $ \Gamma $       & $ 0 $ & $ 5.37 \times 10^{10} $ &  $ 5.7 \times 10^{10} $                             \\
From L        & $ 5.37 \times 10^{10} $ & $ 4.95 \times 10^{10} $ &  $ 5.18 \times 10^{10} $                            \\
From X  & $ 5.7 \times 10^{10} $ & $ 5.18 \times 10^{10} $ &  $ 4.21 \times 10^{10} $                         \\
Intervalley Phonon Energy $ (eV) $     \\
From $ \Gamma $              & $ 0 $ & $ 0.043 $ & $ 0.043 $  \\ 
From L      		& $ 0.043 $ & $ 0.043 $ & $ 0.0411 $  \\
From X 	& $ 0.043 $ & $ 0.041 $ & $ 0.043 $  \\
\hline
\hline
\end{tabular}
\end{table}

\subsection{Scattering Mechanism}

The scattering mechanisms considered in this paper are ionized impurity scattering, polar optical phonon scattering, piezoelectric scattering, acoustic phonon scattering, alloy scattering, non-equivalent, equivalent intervalley scattering and impact ionization scattering. Now we are going to discuss all scattering mechanisms. 

\subsubsection{Ionized Impurity Scattering}
Ionized impurity scattering is an important scattering mechanism at high doping concentrations and at low temperature. Ionized impurity scattering mechanism is considered as an elastic and an anisotropic scattering mechanism. The scattering rate for ionized impurity scattering is given by \cite{Tomizawa}

\begin{equation}
W(E) = \frac{\sqrt{2}e^4 N_I m^{*^{3/2}}}{\pi \epsilon_{s}^2 \hbar^4}(\sqrt{E(1+\alpha E)}(1+2\alpha E)) \left( \frac{1}{q_D^2 \left( q_D^2 + \frac{8m^* E(1+\alpha E) } {\hbar^2} \right)} \right)
\label{ii}
\end{equation}

where $q_D$ is inverse screening length and it is given by

\begin{equation}
q_D = \sqrt{\frac{e^2 N_I}{\epsilon_{s} k_B T} }
\label{inv_deye_length}
\end{equation}

$N_I$ is donor concentration, e is electron charge and $\epsilon_{s} $ is low frequency dielectric constant, $ k_B $ is Boltzmann constant and $ T $ is temperature. 

The angle $\theta$ between initial wave-vector $\boldsymbol{k}$ and final wave-vector $\boldsymbol{k^{'}}$ after ionized impurity scattering, is given by \cite{Ruch_angle}

\begin{equation}
cos \theta = 1 - \frac{2(1-r)}{1-r(\frac{4k^2}{q_D^2})}
\label{angle_ii}
\end{equation}
where $r$ is a uniformly distributed random number between 0 and 1. 

\subsubsection{Polar Optical Phonon Scattering}
Typically polar optical phonon scattering is a dominant scattering mechanism near room temperature and in the higher temperature region. Polar optical phonon scattering is an inelastic and an anisotropic scattering mechanism. The scattering rate for polar optical phonon scattering is given by \cite{Monte2}

\begin{equation}
\begin{split}
W(E) = \frac{e^2 \sqrt{m^*} \omega_{op}}{\sqrt{2} \hbar} \left( \frac{1}{\epsilon_{\infty}} - \frac{1}{\epsilon_s} \right) \frac{1 + 2 \alpha E^{'}}{\sqrt{E ( 1 + \alpha E)}} F_0(E,E^{'}) \times 
\left\{\!\begin{aligned} 
& N_0 & (absorption) \\ 
& (N_0 + 1) & (emission) \\ 
\end{aligned}\right\}
\label{pop}
\end{split}
\end{equation}

where 
\begin{equation}
F_0(E,E^{'}) = C^{-1} \left[ A \hspace{2 mm} ln \left|  \frac{ \sqrt{\gamma(E)} + \sqrt{\gamma(E^{'})} } { \sqrt{\gamma(E)} - \sqrt{\gamma(E^{'})} } \right| + B \right]
\label{F}
\end{equation}

\begin{equation}
A = \left\lbrace  2(1 + \alpha E) (1 + \alpha E^{'}) + \alpha [\gamma(E) + \gamma(E{'})] \right\rbrace^2
\label{A}
\end{equation}

\begin{equation}
B = - 2 \alpha \sqrt{\gamma(E) \gamma(E^{'})} \left[ 4 (1 + \alpha E) (1 + \alpha E^{'}) + \alpha \left\lbrace  \gamma(E) + \gamma(E^{'})\right\rbrace \right]
\label{B}
\end{equation}

\begin{equation}
C = 4 (1 + \alpha E) (1 + \alpha E^{'} )(1 + 2 \alpha E) (1 + 2 \alpha E^{'})
\label{C}
\end{equation}

where $\epsilon_{\infty}$ is high frequency dielectric constant, $\omega_{op}$ is polar optical phonon frequency, $E^{'} = E + \hbar \omega_{op} $ for absorption and $E^{'} = E - \hbar \omega_{op} $ for emission of polar optical phonon, if $E^{'} < 0$ polar optical phonon scattering will not occur, $N_0$ is the number of phonons involved in the transition. $N_0$ is given by   

\begin{equation}
N_0 = \frac{1}{e^{\frac{\hbar \omega_{op}}{k_B T}} - 1}
\label{omega_pop}
\end{equation}

The angle $\theta$ between initial wave-vector $\boldsymbol{k}$ and final wave-vector $\boldsymbol{k^{'}}$ for polar optical phonon scattering, is given by the following probability distribution function \cite{GaAs,CdTe_2}
\begin{equation}
P(cos\theta) d(cos\theta) = a_{pop} \frac{ (\sqrt{\gamma(E) \gamma(E^{'})} + \alpha E E^{'} cos \theta)^{2}}{\gamma(E) + \gamma(E^{'}) - 2 \sqrt{ \gamma(E) \gamma(E^{'}) } cos \theta} d(cos \theta)
\label{angle_pop}
\end{equation}

where $a_{pop}$ is a normalization constant.
The random values of $cos\theta$ with the above probability distribution is obtained by using Von Neumann rejection technique \cite{GaAs,Neumann}.  

\subsubsection{Piezoelectric Scattering}
Piezoelectric scattering is an important scattering mechanism at low doping density and low temperature in polar materials. Piezoelectric scattering is treated here by elastic and equipartition approximation. The piezoelectric scattering rate is given by \cite{Vasileska, Ridley} 

\begin{equation}
W(E) = \frac{m^{*^{1/2}} e^2 P_{pz}^2 k_B T}  {4 \sqrt{2} \pi \hbar^2 \epsilon_{s}} 
\left( \frac{ 1 + 2 \alpha E }{ \sqrt{ E (1 + \alpha E) } } \right) ln \left( 1 + \frac{ 8 m^* E (1 + \alpha E )}{\hbar^2 q_D^2} \right)
\label{pz}
\end{equation}

where $P_{pz}$ is dimensionless piezoelectric coefficient.

The angle $\theta$ between initial wave-vector $\boldsymbol{k}$ and final wave-vector $\boldsymbol{k^{'}}$ for piezoelectric scattering, is given by the following equation
\cite{Monte_thesis}

\begin{equation}
cos\theta = 1 + \frac{\hbar^2 q_D^2}{4m^* \gamma(E)} \left[ 1 - \left( 1 + \frac{8m^*\gamma(E)}{\hbar^2 q_D^2}\right)^r \right]
\label{angle_pz}
\end{equation}

where r in an uniformly distributed random number between 0 and 1.

\subsubsection{Acoustic Phonon Scattering}
Acoustic phonon scattering occur due to scattering of electrons by non polar acoustic phonons. Acoustic phonon scattering is treated by elastic and equipartition approximation. Acoustic phonon scattering is given by \cite{Jacoboni-1-review,Tomizawa}
      
\begin{equation}
W(E) = \frac{\sqrt{2}m^{*^{3/2}} k_B T D_{ac}^2}{\pi \hbar^4 \rho v_s^2} \sqrt{E(1+\alpha E)}(1+2\alpha E)
\label{ac}
\end{equation}
where $D_{ac}$ is acoustic deformation potential, $\rho$ is density of material and $v_s$ is sound velocity.

The angle $\theta$ between initial wave-vector $\boldsymbol{k}$ and final wave-vector $\boldsymbol{k^{'}}$ for acoustic phonon scattering, is given by the following probability distribution function \cite{GaAs,CdTe_2}

\begin{equation}
P(cos\theta) d(cos\theta) = a_{ac} \{ 1 +  \alpha E \left( 1 + cos \theta \right) \} ^2 d(cos \theta)
\label{angle_ac}
\end{equation}

where $a_{ac}$ is a normalization constant.
The random values of $cos\theta$ with the above probability distribution is obtained by using Von Neumann rejection technique \cite{GaAs,Neumann}.

\subsubsection{Alloy Scattering}
In semiconductor alloys, there is one more additional scattering mechanism of free carriers occur due to random fluctuations of perfect periodicity of the crystal. The alloy scattering rate for electrons is given by \cite{alloy1,alloy2,alloy3}  

\begin{equation}
W(E) = \frac{3 \pi}{8 \sqrt{2}} \frac{m^{*{\frac{3}{2}}}}{\hbar^4} x (1-x) V_0 U_{all}^2 (1+2 \alpha E) S(E) \sqrt{E(1+\alpha E)}
\label{alloy}
\end{equation}

where $x$ is mole fraction, $V_0$ is the primitive cell volume and $ U_{all} $ is alloy scattering potential. We have taken a value of $ 0.47 eV $ for alloy scattering potential \cite{AlInAs_paper1}. Here, $ S(E) $ is an energy-dependent parameter that describe the effect of alloy ordering on the scattering rate. Value of $S(E)$ lies between $0$ and $1$. $S(E) = 0$ refers perfectly ordered alloy system and $S(E) = 1$ refers to completely random alloy system. Throughout the simulation, $S = 1$ is considered. Alloy scattering is an isotropic scattering mechanism and it is treated by using elastic approximation.   

\subsubsection{Intervalley Phonon Scattering}
The scattering rate due to intervalley phonon is given by \cite{Jacoboni-1-review, Tomizawa}

\begin{equation}
W(E) = \frac{\pi D_i^2 Z}{\rho \omega_i} \left( \frac{(2m^*)^{\frac{3}{2}} \sqrt{E^{'}(1+ \alpha E^{'})} (1 + 2 \alpha E^{'}) } {4 \pi^2 \hbar^3 } \right) \times
\left\{\! \begin{aligned} 
& N(\omega_i) & (absorption) \\
& (N(\omega_i) + 1) & (emission) \\ 
\end{aligned} \right\} 
\label{iv}
\end{equation}

where $E^{'} = E + \hbar \omega_i - \triangle E$ for absorption and $E^{'} = E - \hbar \omega_i - \triangle E$ for emission of intervalley phonon, if $E^{'} < 0$ intervalley scattering will not occur. For intra-valley scattering $\triangle E = 0$ and for intervalley scattering $\triangle E $ is the difference between bottom of energy band between two valleys. $D_i$ is intervalley scattering coupling constant, $Z$ is the number of final valley for intervalley scattering, $N(\omega_i)$ is the number of phonons involved in the transition, $\omega_i$ is intervalley phonon frequency. $N(\omega_i)$ is given by   

\begin{equation}
N(\omega_i) = \frac{1}{e^{\frac{\hbar \omega_i}{k_B T}} - 1}
\label{omega_i}
\end{equation}

Intervalley scattering is considered here to be isotropic in nature. So, final state after intervalley are equally probable, restricted to only conservation of energy. 

\subsubsection{Impact Ionization Scattering}
The scattering rate due to impact ionization is treated by using Keldysh expression \cite{qq}
  
\begin{equation}
 \frac{1}{\tau_{ii}(E)} = \left\{
						\begin{array}{ll}
						 0   \quad  \quad \quad \quad  \quad \quad \quad   \quad E < E_{th} \\	 
						 \frac{P}{\tau(E_{th})}\left(\frac{E - E_{th}}{E_{th}}\right)^2    \quad   \quad    E > E_{th} \\
						 \end{array}
						 \right\}. 
\label{impact}	
\end{equation}

where $\frac{1}{\tau_{ii}(E)}$ is impact ionization scattering rate for an electron. $\frac{1}{\tau(E_{th})}$ is the scattering rate at the threshold energy $E_{th}$ and $ P $ is a dimensional less coupling constant. In our simulation Threshold energy $E_{th}$ and $ P $ is treated as fitting parameters.     

The angle $\theta$ between initial wave-vector $\boldsymbol{k}$ and final wave-vector $\boldsymbol{k^{'}}$ for impact ionization scattering, is given by the following equation \cite{Curby}

\begin{equation}
cos\theta = - \frac{1+G(1-2r)}{2r- (G + 3)}
\label{angle_impact}
\end{equation}

where  

\begin{equation}
G = \frac{k^2 + k^{'2} + \gamma^2}{k k^{'}}
\label{impact_const}
\end{equation}

For all scattering mechanisms azimuthal angle $\phi $ is completely random, so $\phi$ can be easily calculated by using a uniformly distributed random number r between 0 and 1, by $\phi = 2 \pi r$. The magnitude of final state wave-vector $\boldsymbol{k^{'}}$, is determined by using energy conservation for the given scattering mechanism.  

\subsection{Diffusion}
Diffusion constant is the one of the important parameter to understand the carrier transport in semiconductors. In the recent past, a lot of work has been done with Monte Carlo technique for high field diffusivity calculation specially in small semiconductor devices \cite{diff1,diff2,diff3,diff4}. At lower field Diffusion $D$ and mobility $\mu $ are related by Einstein relation $D = \frac{\mu k_B T}{e}$. At higher electric field or in the presence of intervalley scattering, Einstein relation fails and  diffusion constant can not be calculated by using the Einstein relation. In the present work, the following equation is used to calculate the diffusion constant along the longitudinal direction \cite{diff3,diff5}

\begin{equation}
D_l = \frac{<(x_l(t)-<x_l(t)>)^2>}{2t}
\label{Dl}
\end{equation}

where $x_l(t)$ is the displacement along external field direction at time $t$, and the brackets $<...>$ denotes the ensemble averages. While doing evaluation with Eq. \ref{Dl} both ensemble and time averages are taken into account in Monte Carlo simulation. For evaluation of diffusivity along the transverse direction to the field $D_t$ same expression as \ref{Dl} is used just by replacing displacement along the parallel direction to the field with transverse direction to the electric field. The above Eq. \ref{Dl} neglects the electron-electron repulsion and assumes that electric field is constant everywhere and the Eq. \ref{Dl} is valid only when macroscopic Fick's law  is applicable. The Eq. \ref{Dl} is obtained from the Fick's law given below,
\begin{equation}
\frac{\partial n}{\partial t}=D_l\frac{\partial^2 n}{\partial x_l^2}-v_d\frac{\partial n}{\partial x_l}
\end{equation}
Here $n$ being the electron density and $v_d$ is the drift velocity. Equation \ref{Dl} is obtained from the second moment of the electron density. For the  transient conditions, as Fick's law does not hold, Eq. \ref{Dl} can not be employed. For the present study, we report the diffusivity only for the steady state conditions.   

\subsection{Relaxation Time}

The momentum relaxation time $\tau_m$ at steady state is calculated by using the following equation \cite{diff3,rel2,rel3,rel4}

\begin{equation}
\tau_m = \frac{m_{eff} v_{ss}}{q F}
\label{rel_momentum}
\end{equation}
where $m_{eff}$ is effective mass over ensemble and is given by $m_{eff} = m^*(1+2\alpha E)$. $v_{ss}$ is steady state average drift velocity of electrons and $F$ is applied electric field.

The energy relaxation time $\tau_e$ at steady state is calculated by using the following equation \cite{diff3,rel2,rel3,rel4}
\begin{equation}
\tau_e = \frac{\overline{E}-\overline{E}_0}{e v_{ss} F}
\label{rel_energy}
\end{equation}  

where $\overline{E}$ is the average electron energy in the presence of field and $\overline{E}_0 = \frac{3}{2}k_B T$ is the thermal energy.
The energy relaxation time will give insights into the transient effect of the material. Generally higher energy relaxation time leads to higher transient effects \cite{rel4,seeger}.

\section{Results and Discussion}

\subsubsection{Steady State Electron Transport}

\begin{figure}
\includegraphics[height=0.37\textwidth,width=0.48\textwidth]{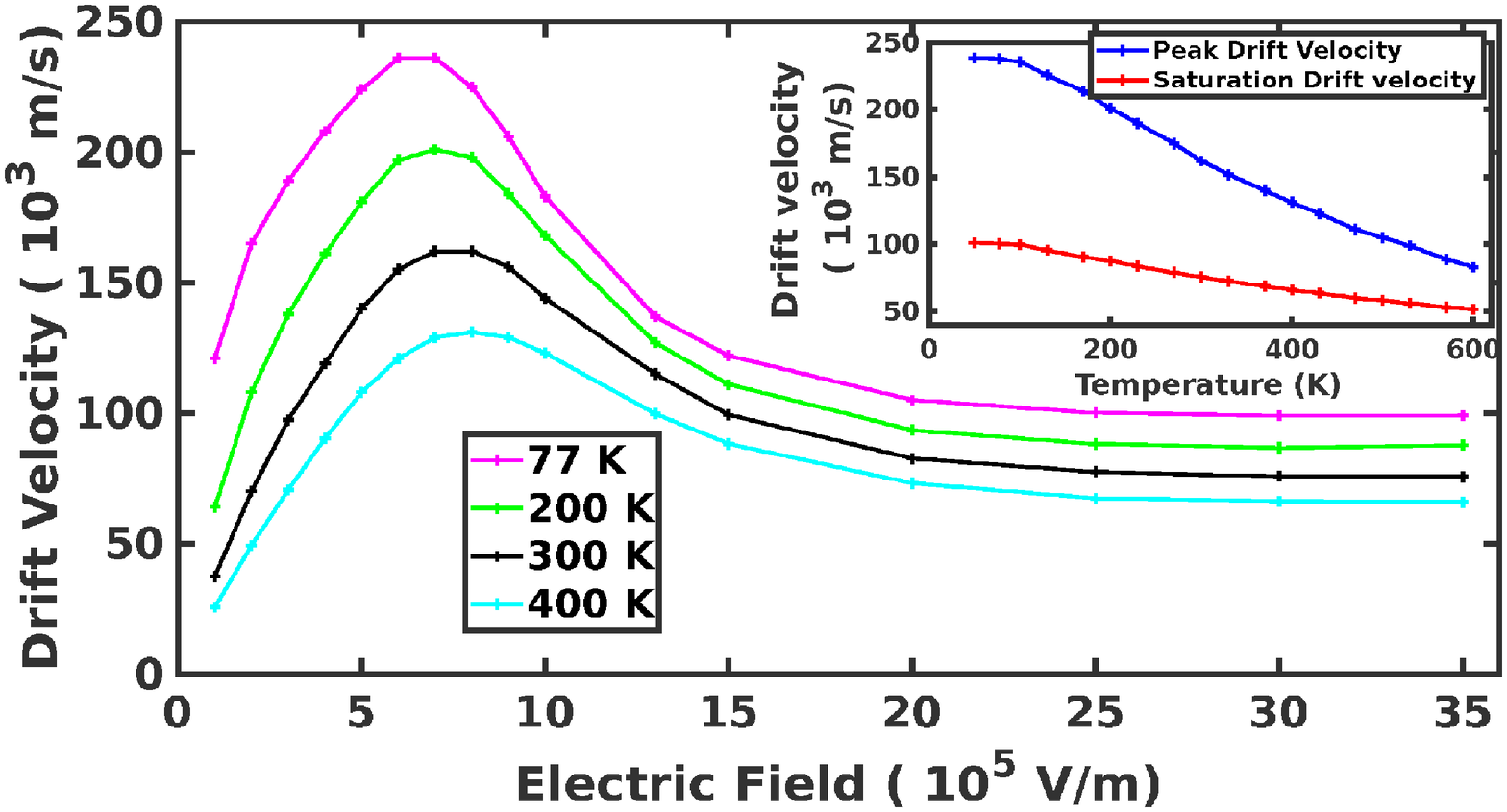} 
\caption{Drift velocity as a function of applied  electric field for different temperatures }
\label{fig1}
\end{figure}

\begin{figure}
\subfigure[$ $]{\includegraphics[height=0.37\textwidth,width=0.48\textwidth]{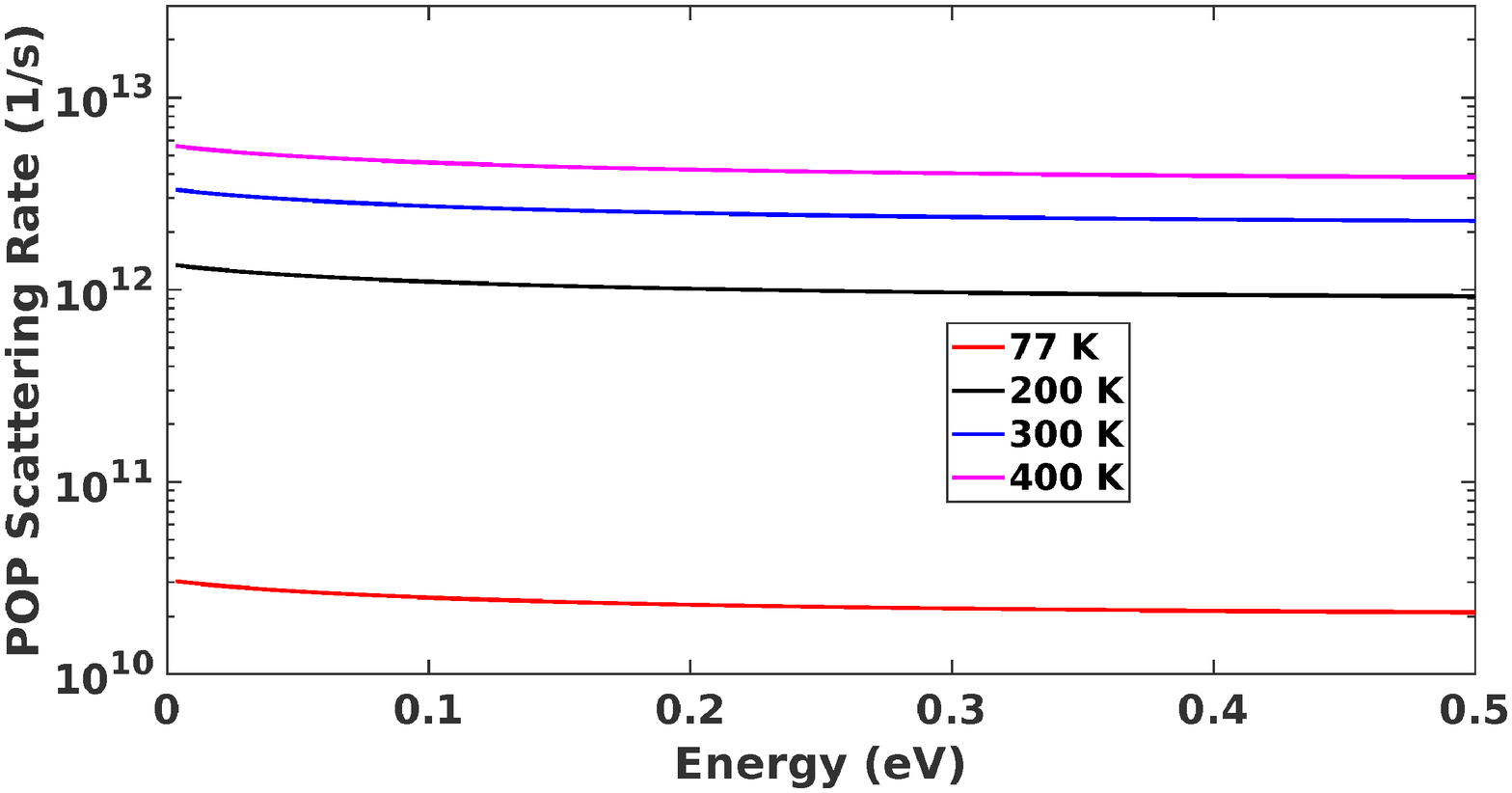} }
\subfigure[$ $]{\includegraphics[height=0.37\textwidth,width=0.48\textwidth]{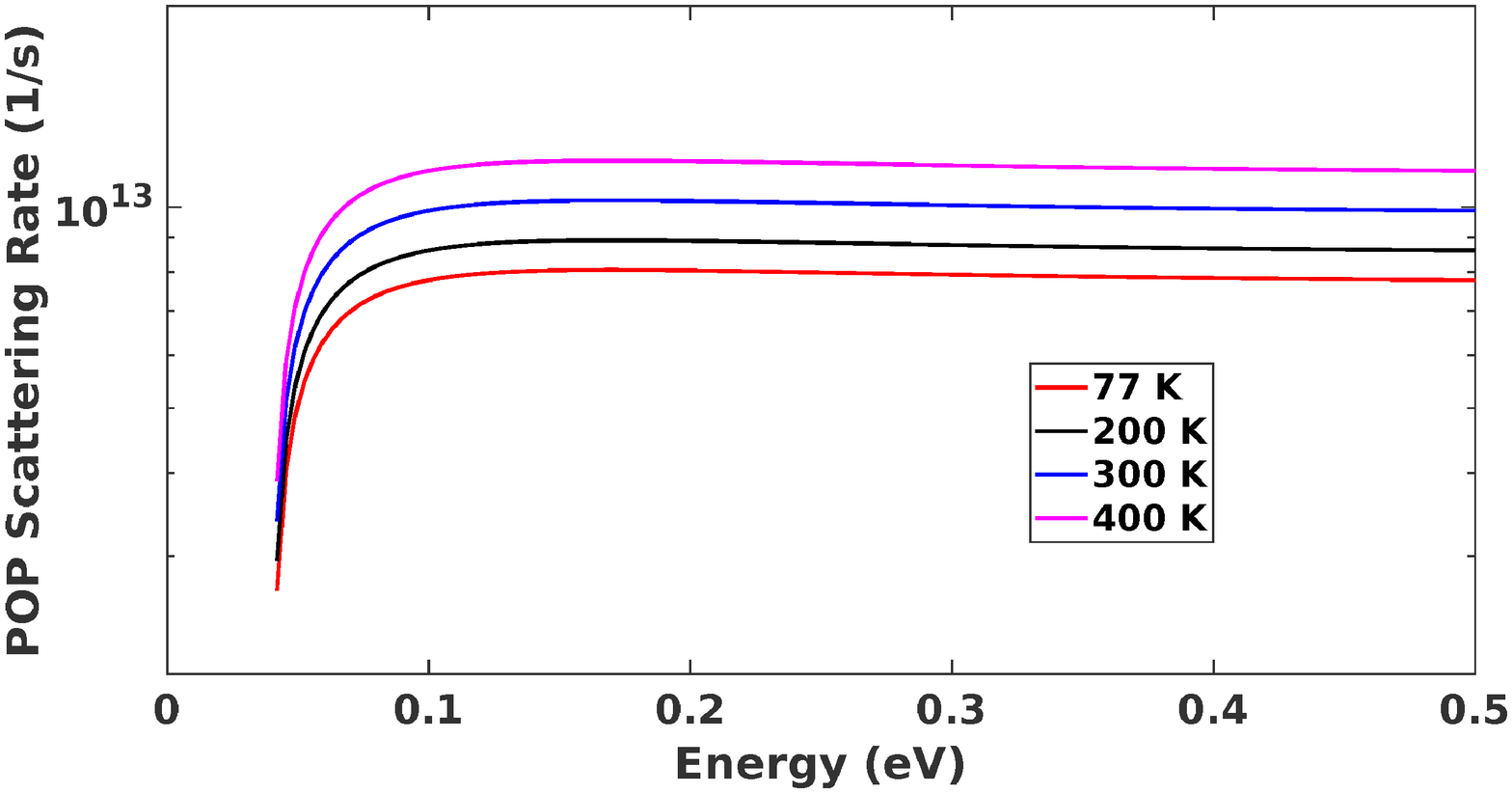} }
\caption{Polar optical phonon scattering rate as a function of electron energy for different temperature in the gamma valley due to (a) Absorption of optical phonon (b) Emission of optical phonon}
\label{pop_scat}
\end{figure}

\begin{figure}
\subfigure[$ $]{\includegraphics[height=0.37\textwidth,width=0.48\textwidth]{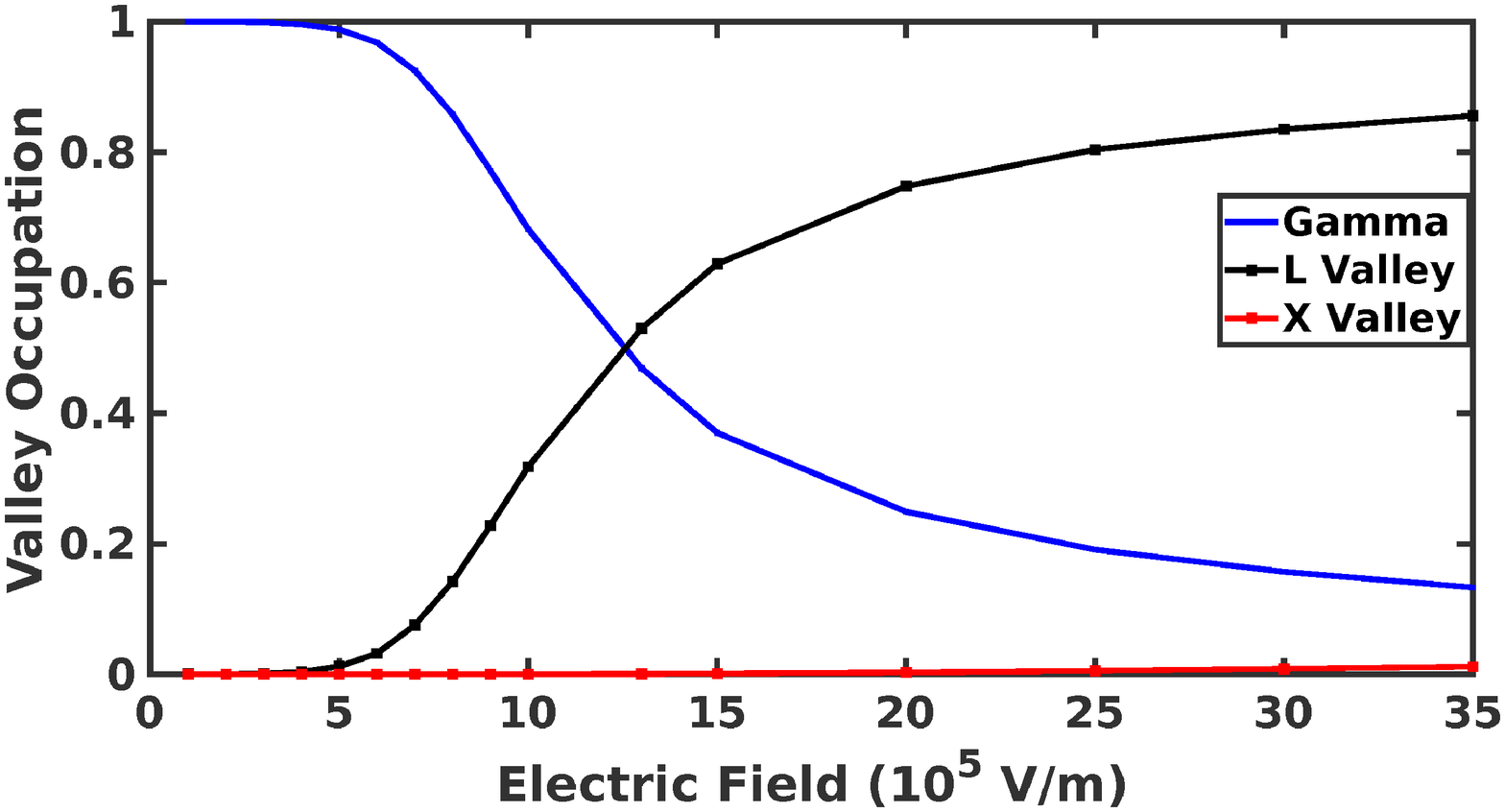} }
\subfigure[$ $]{\includegraphics[height=0.37\textwidth,width=0.48\textwidth]{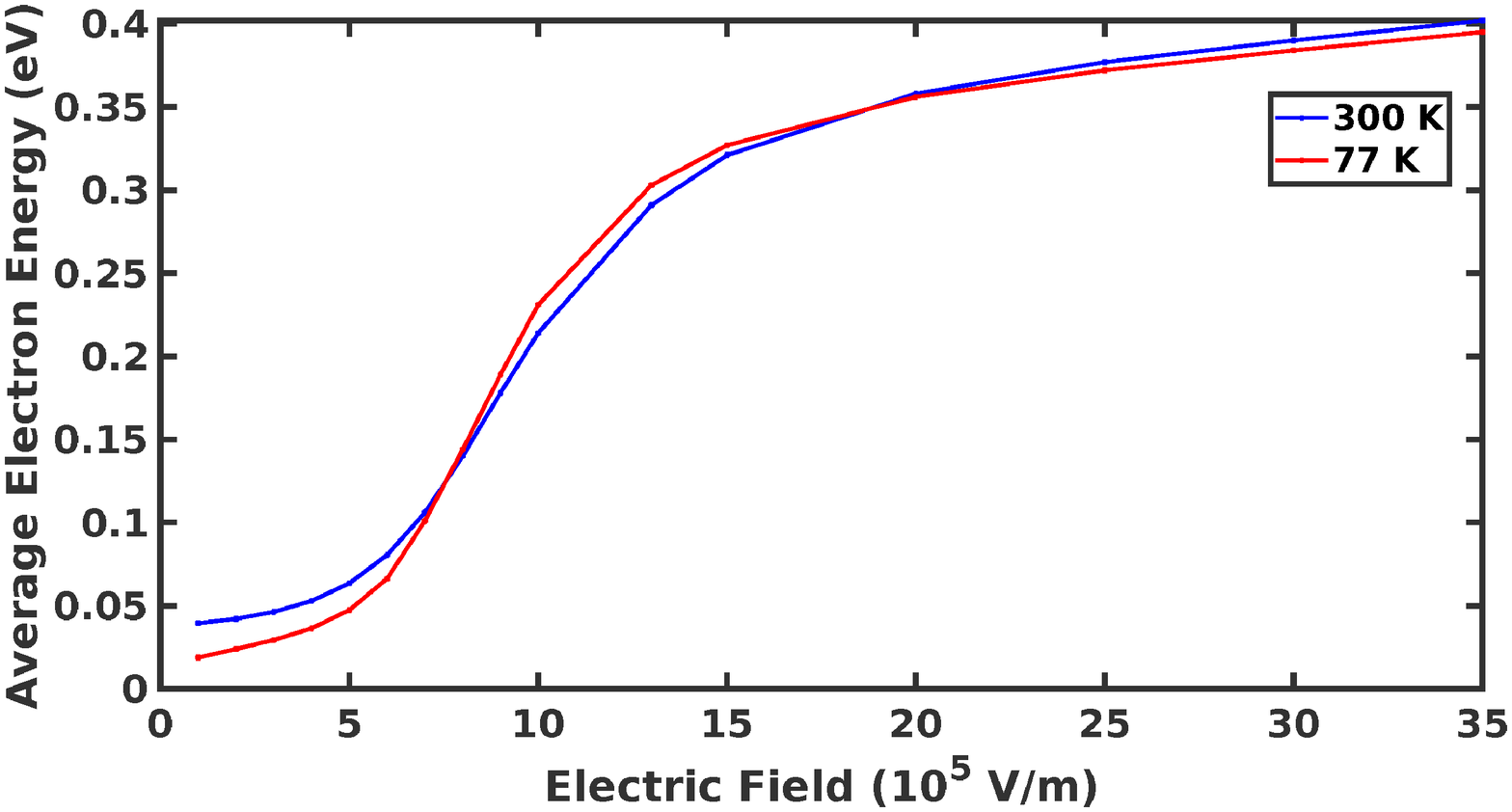} }
\caption{(a) Relative population in different valleys as a function of electric field at $300$ K (b) Average total electron energy as a function of electric field at $77$ K and $300$ K}
\label{fig2}
\end{figure}

In the Fig. \ref{fig1} we show the drift velocity with electric field for different lattice temperatures. It can be seen that both the peak and the saturation velocity show significant  temperature dependence. This can further be noticed from the inset figure in Fig. \ref{fig1} where peak and saturation drift velocity are plotted with respect to temperature. To understand such temperature dependence, we plot the scattering rates due to the absorption (Fig. \ref{pop_scat} (a)) and due to the emission of polar optical phonons (Fig. \ref{pop_scat} (b)). It can be seen that the electron scattering through the emission of polar optical phonons takes the dominant role in bringing the temperature dependence in the drift velocity. Another reason of such temperature dependence is the small energy difference between the $\Gamma$ and  L valley ($\sim$ 340 meV). It can be seen from the Fig. \ref{fig2} (a) that at T = 300 K, for electric field as low as 7.5$\times$10$^5$ V/m, the L valley is populated with about the 20\% of the total electrons. The drift velocity then starts reducing due to (i) inter-valley scattering (ii) higher effective mass of the L valley. This particular value of the electric field can be called as threshold or critical field.

The Fig. \ref{fig2} (b) shows the average total electron energy as a function of electric field at $77$ K and $300$ K. Near threshold field, there is significant sharp increase in electron energy with electric field. Above an electric field strength of $20 \times 10^5 $ V/m the electron energy gets saturated and there is very slow increase in the electron energy. Near threshold field, most of the electrons are in the gamma valley and polar optical phonon scattering by emission of phonons is most dominant scattering mechanism and it relaxes energy of electron by only $0.0397 $ eV, while at higher electric field above $20 \times 10^5 $ V/m, most of the electrons are shifted to higher valley and intervalley scattering becomes dominant scattering mechanism, second polar optical scattering in L-valley also have higher scattering rate than polar optical phonon scattering of gamma valley. So, electron energy shows a sharp upward turn near threshold field, while at higher field it shows small variation with the electric field.

\begin{figure}
\subfigure[$ $]{\includegraphics[height=0.37\textwidth,width=0.48\textwidth]{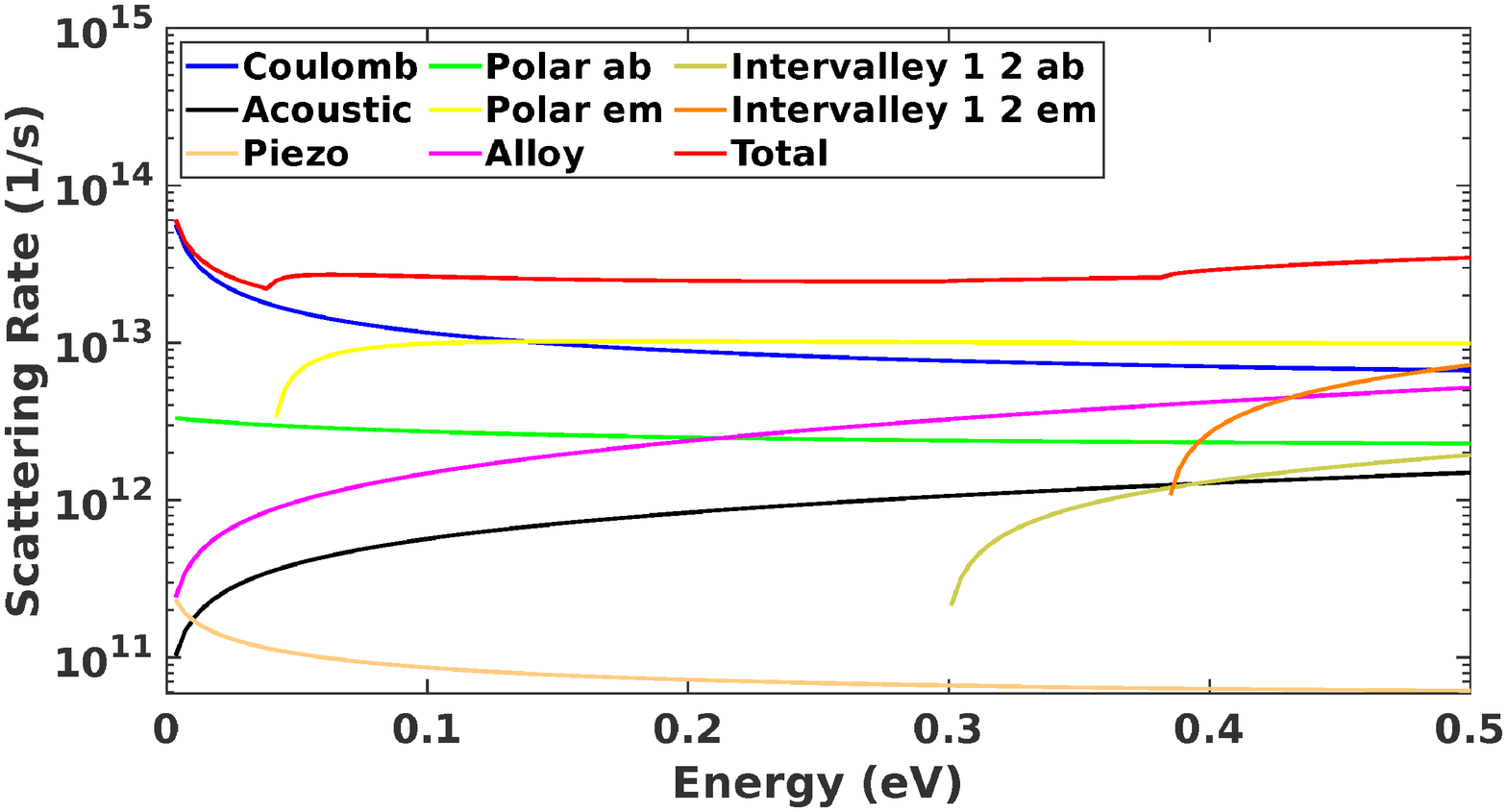} }
\subfigure[$ $]{\includegraphics[height=0.37\textwidth,width=0.48\textwidth]{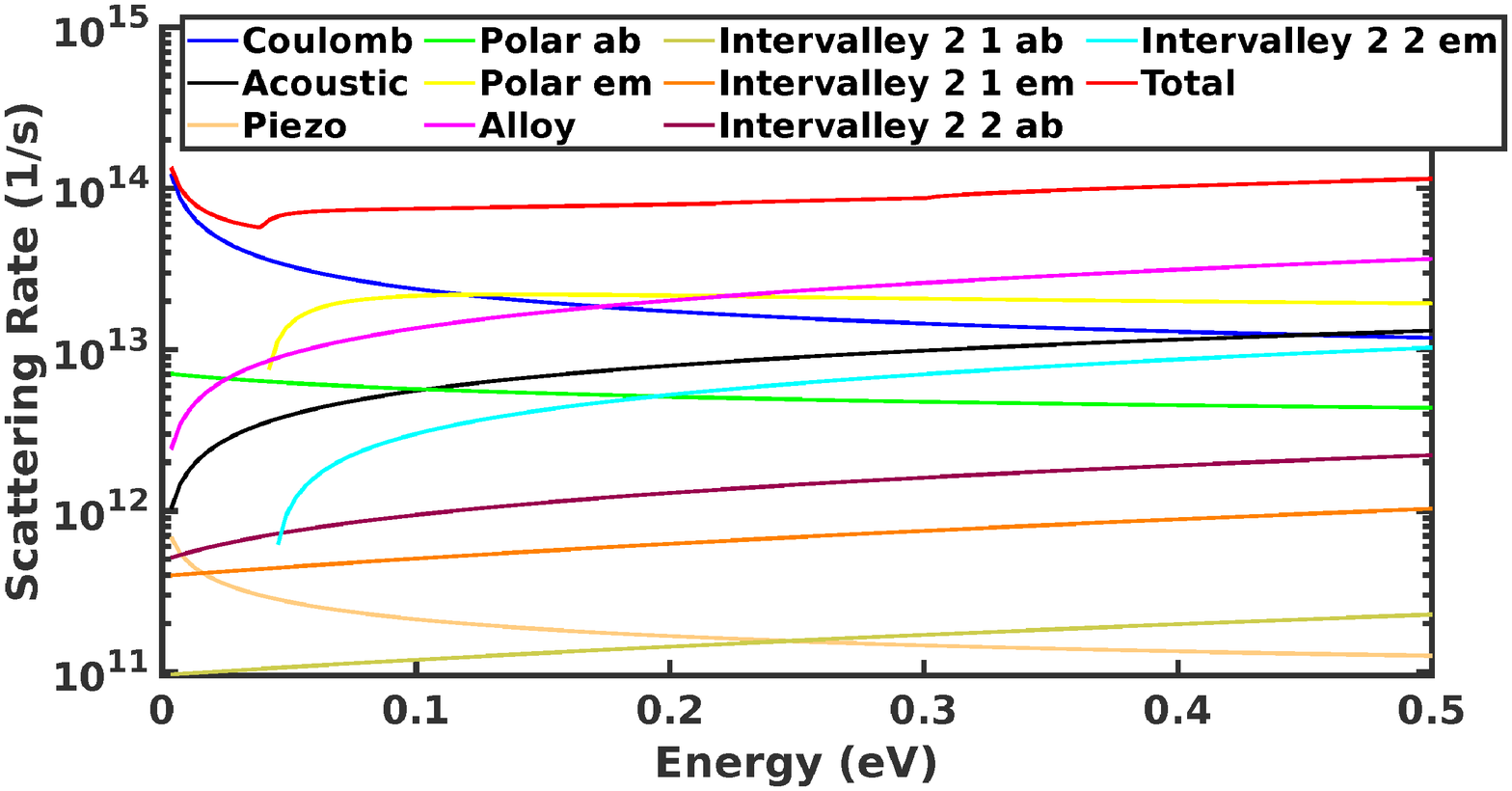} }
\caption{(a) Scattering rate in gamma valley as a function of energy at $300$ K (b) Scattering rate in L valley as a function of energy at $300$ K}
\label{fig3}
\end{figure}

\begin{figure}
\includegraphics[height=0.37\textwidth,width=0.48\textwidth]{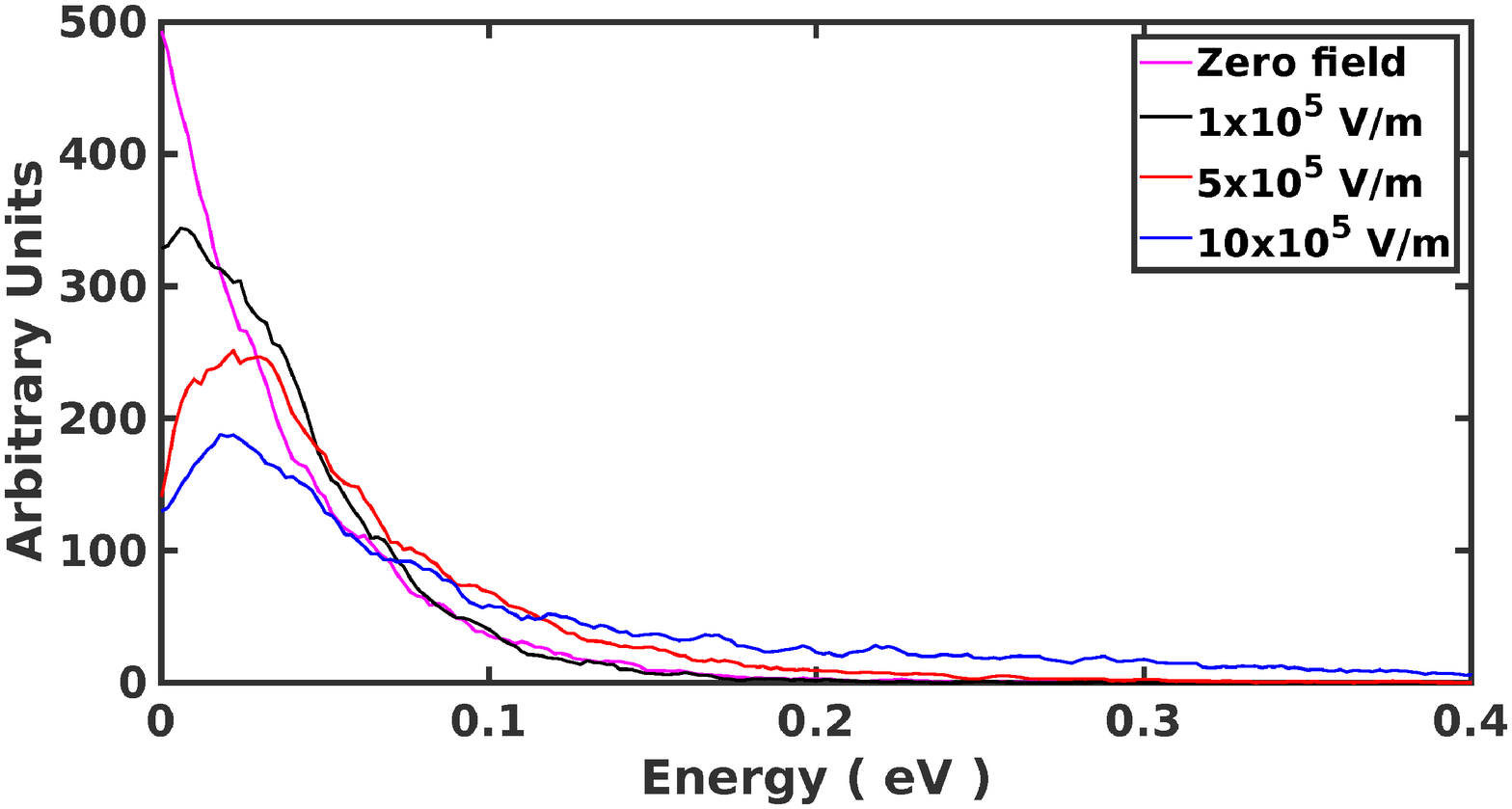} 
\caption{The electron energy distribution function for different applied electric field} 
\label{energy_dist}
\end{figure}

\begin{figure}
\includegraphics[height=0.37\textwidth,width=0.48\textwidth]{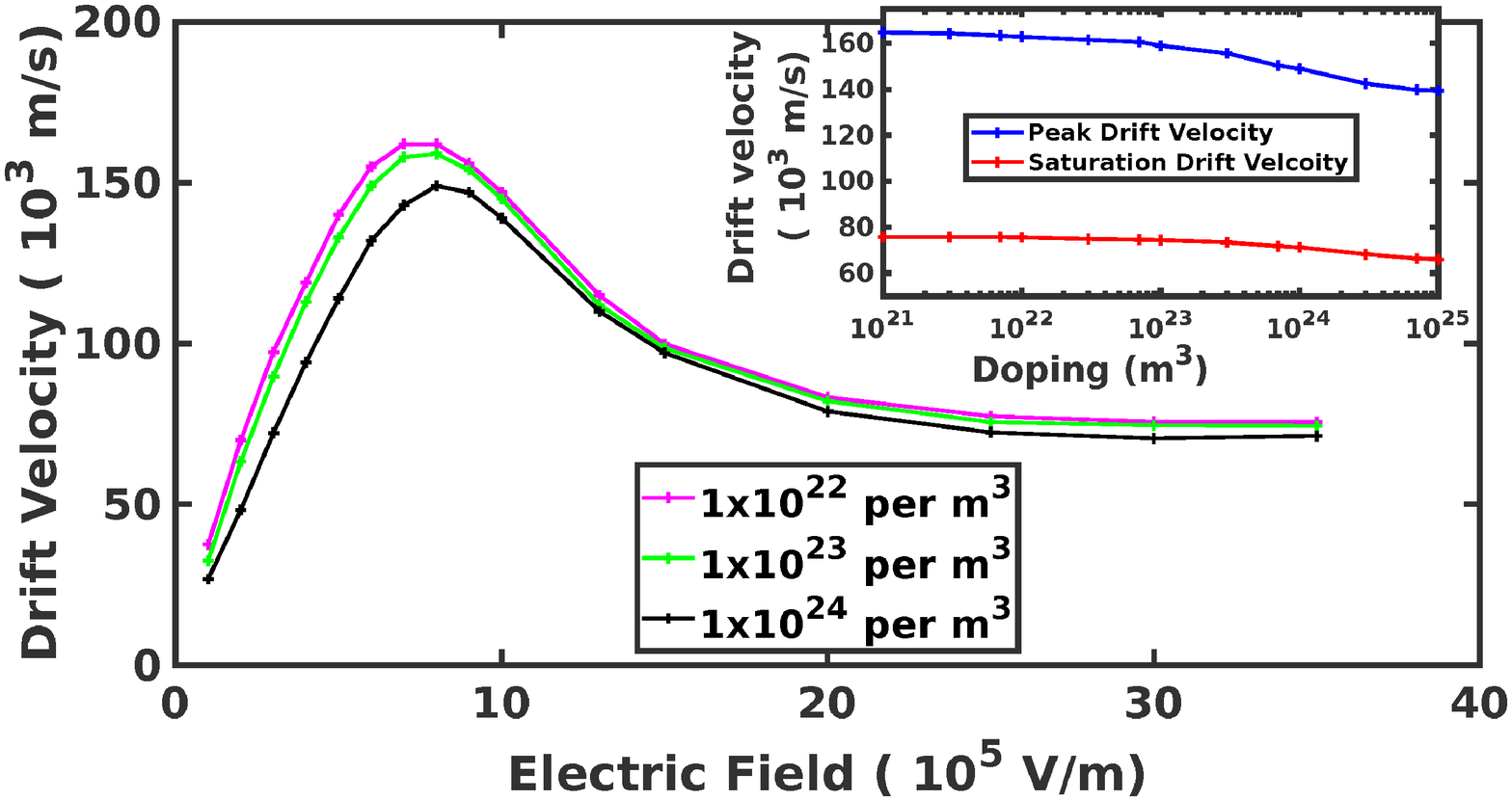} 
\caption{Drift velocity as a function of applied  electric field for different doping concentrations at $300$ K }
\label{fig8}
\end{figure}

In Fig. \ref{fig3} (a) the scattering rate for different scattering mechanisms and total scattering rate except impact ionization scattering rate as a function of electron energy are plotted for the central gamma valley at $ 300$ K. At low electron energy ionized impurity scattering is most dominant scattering and above approximately $0.14$ eV polar optical phonon scattering by emission of phonon is the most dominant scattering mechanism. 
    
In Fig. \ref{fig3} (b) the scattering rate for different scattering mechanisms and total scattering rate as a function of energy are plotted for the L-valley at $300$ K. Since, most of the electrons remains in gamma and L valley for the electric field of interest, so intervalley scattering of carriers to X valley is not shown for convenience in the Figs. \ref{fig3} (a) and \ref{fig3} (b), but it is included in our simulation. The total scattering rate in the L-valley is higher than the total scattering rate in the gamma valley, this is because of higher density of states in the L-valley due to the higher effective mass of electrons there.   Figure \ref{energy_dist} shows the variation of electron distribution function for different applied electric field strengths. As the applied field strength increases distribution of carriers at higher energy region increases. 

Figure \ref{fig8} shows the variation of drift velocity with electric field for different doping concentrations. As doping concentration increases drift velocity, peak velocity and low field mobility get reduced and threshold field is shifted to higher electric field values.	With the increase in doping concentration, ionized impurity scattering rate increases since the ionized impurity scattering rate is directly proportional to doping concentration, so drift velocity and peak velocity reduces. Higher ionized impurity scattering rate at higher doping concentration causes lower electron energy and increases electric field needed to reach peak drift velocity. From Fig. \ref{fig3} (a) it is clear that at lower electron energy ionized impurity scattering is most dominant scattering mechanism, while at higher electron energy ionized impurity scattering is not so significant. Our simulation results in Fig. \ref{fig8} also depicts the same, at lower field ionized impurity scattering has significant effect while at higher electric field saturation velocity is not significantly affected by doping concentration variation. The inset figure of Fig. \ref{fig8} shows the variation of peak drift velocity and saturation drift velocity with doping concentration.

\begin{figure}
\subfigure[$ $]{\includegraphics[height=0.37\textwidth,width=0.48\textwidth]{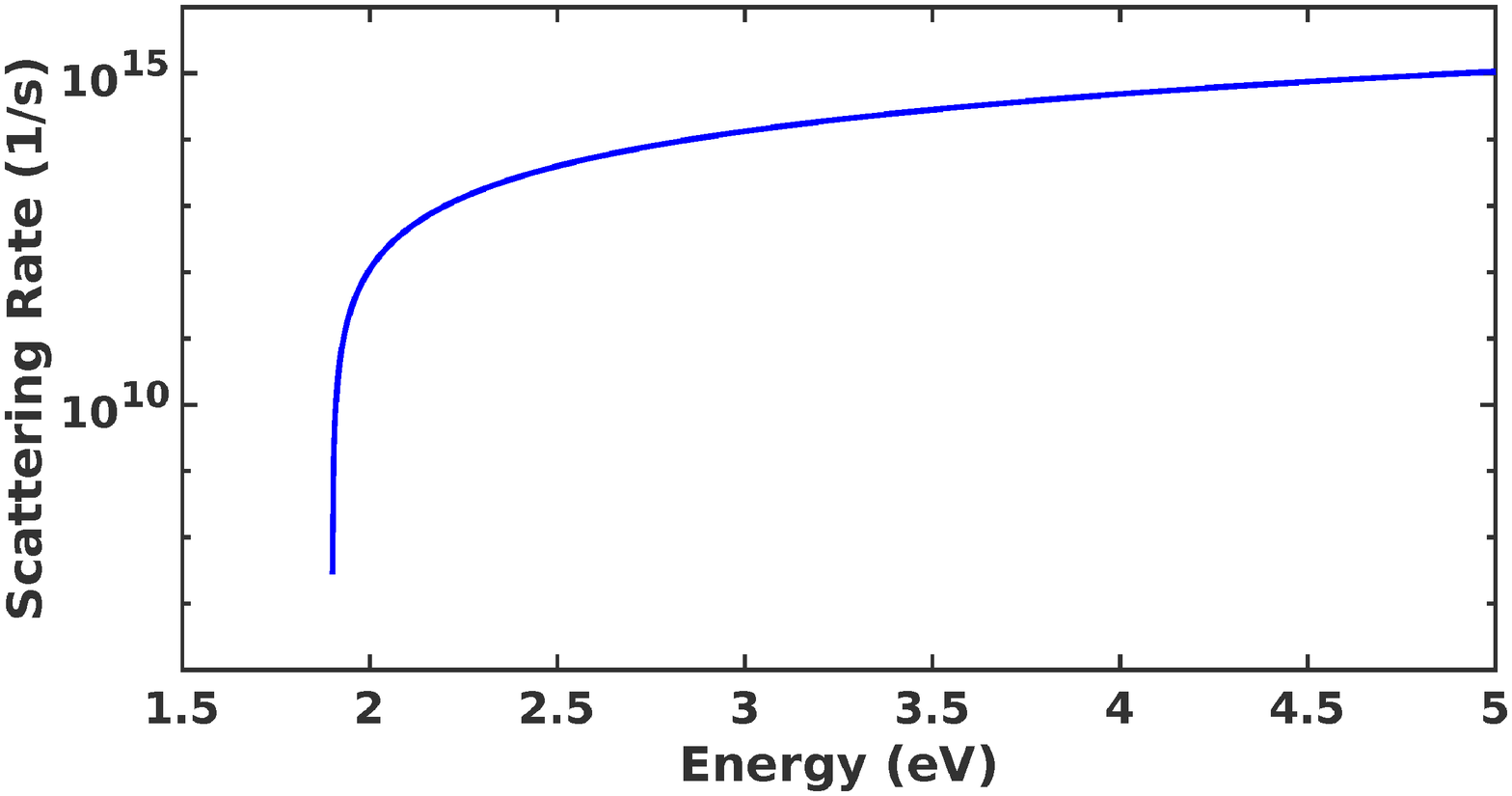} }
\subfigure[$ $]{\includegraphics[height=0.37\textwidth,width=0.48\textwidth]{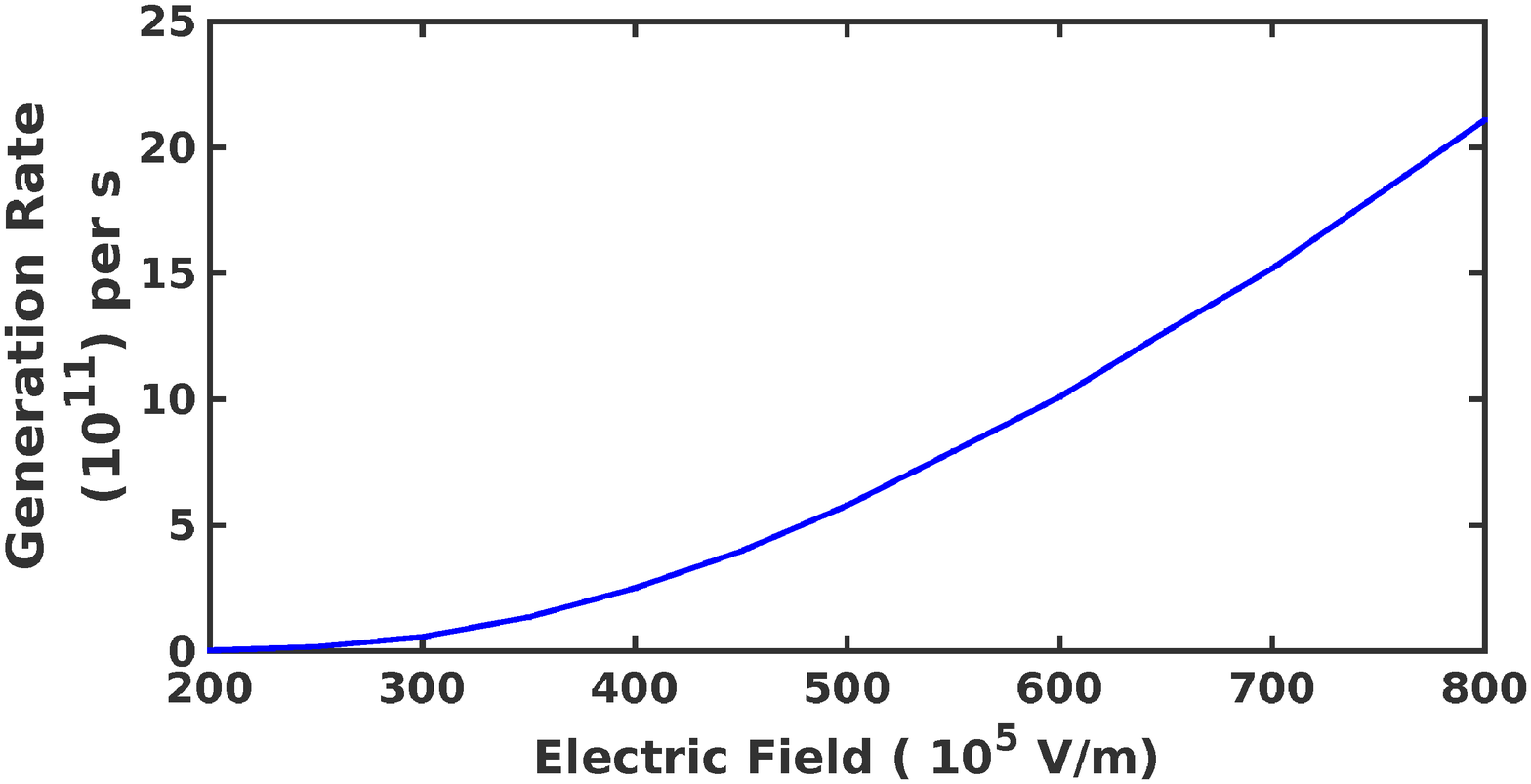} }
\subfigure[$ $]{\includegraphics[height=0.37\textwidth,width=0.48\textwidth]{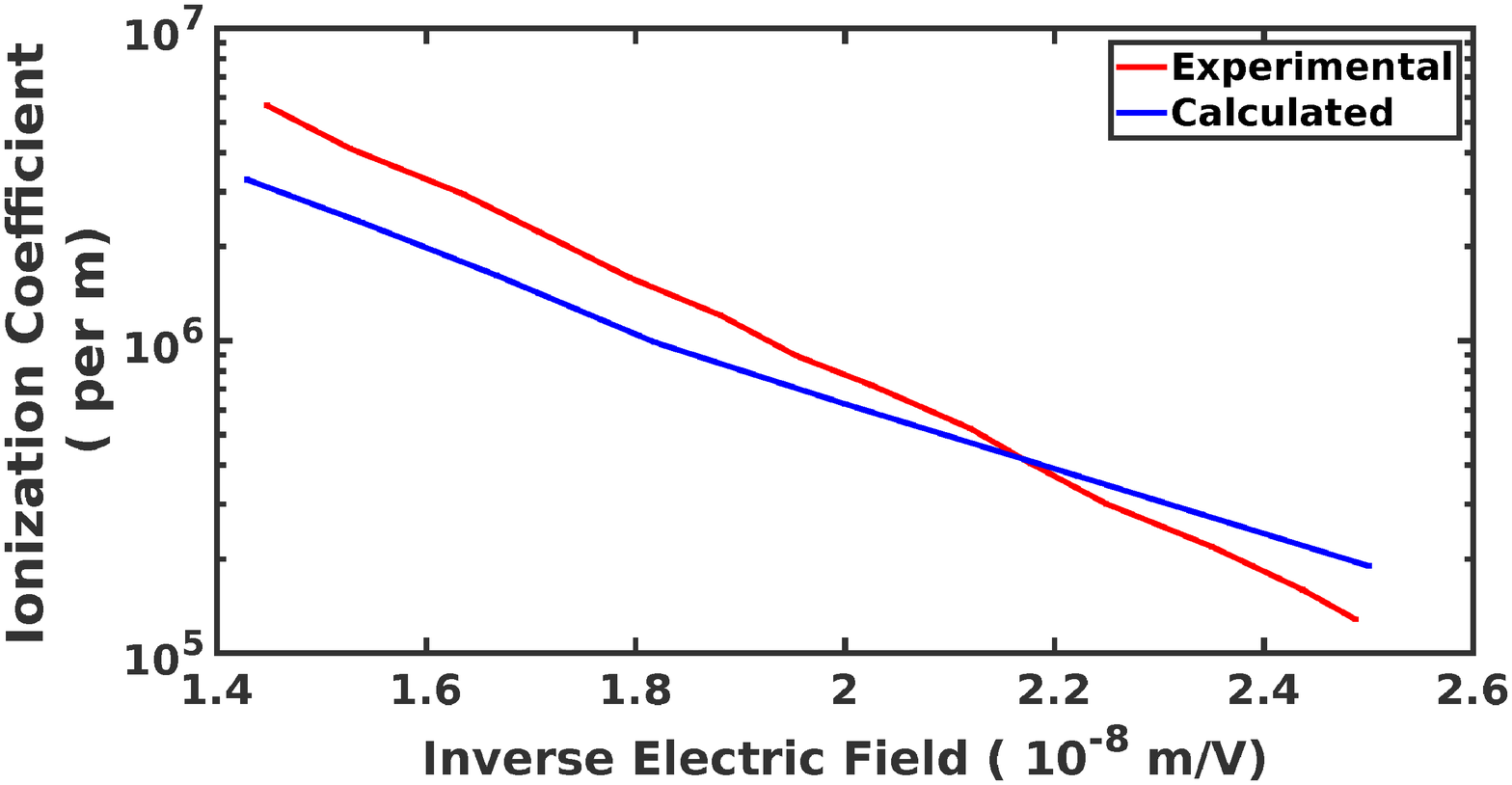} }
\caption{(a) Impact ionization scattering rate in gamma valley as a function of energy. (b) Impact ionization generation rate as a function of applied electric field (c) Impact ionization coefficient as a function of applied electric field. For all cases doping concentration is set to $2 \times 10^{22} m^{-3}$ and crystal temperature is $298$ K }
\label{fig5}
\end{figure}

Figure \ref{fig5} (a) shows the scattering rate due to impact ionization scattering. The threshold energy of impact ionization in gamma valley is $ 1.9 $ eV, so impact ionization becomes active only after $1.9$ eV energy in gamma valley. Figure \ref{fig5} (b) shows the variation of the generation rate due to impact ionization with electric field obtained in our simulation for doping concentration of $2 \times 10^{22} m^{-3} $ at $298$ K. For impact ionization threshold energy $E_{th}$ and $P$ is treated as fitting parameter, and their value we have obtained are written in table \ref{table3} for different conduction bands. A good agreement between the experimental and theoretical curve of impact ionization coefficient with inverse electric field is obtained as shown in Fig. \ref{fig5} (c).

\begin{table}
\caption{ Parameters for impact ionization }
\label{table3}
\begin{ruledtabular}
\begin{tabular}{cccccccc}
Parameter  &  $E_{th} (eV) $  &  $ P $    \\
\hline
First conduction band      & 1.9  &  $4 \times 10^{14} $ \\
Second conduction band     & 2.3  &  $1 \times 10^{15} $ \\
Third conduction band      & 2.4  &  $1 \times 10^{16} $ \\
\end{tabular}
\end{ruledtabular}
\end{table}   


\subsubsection{Momentum and Energy Relaxation Time}

\begin{figure}
\subfigure[$ $]{\includegraphics[height=0.37\textwidth,width=0.48\textwidth]{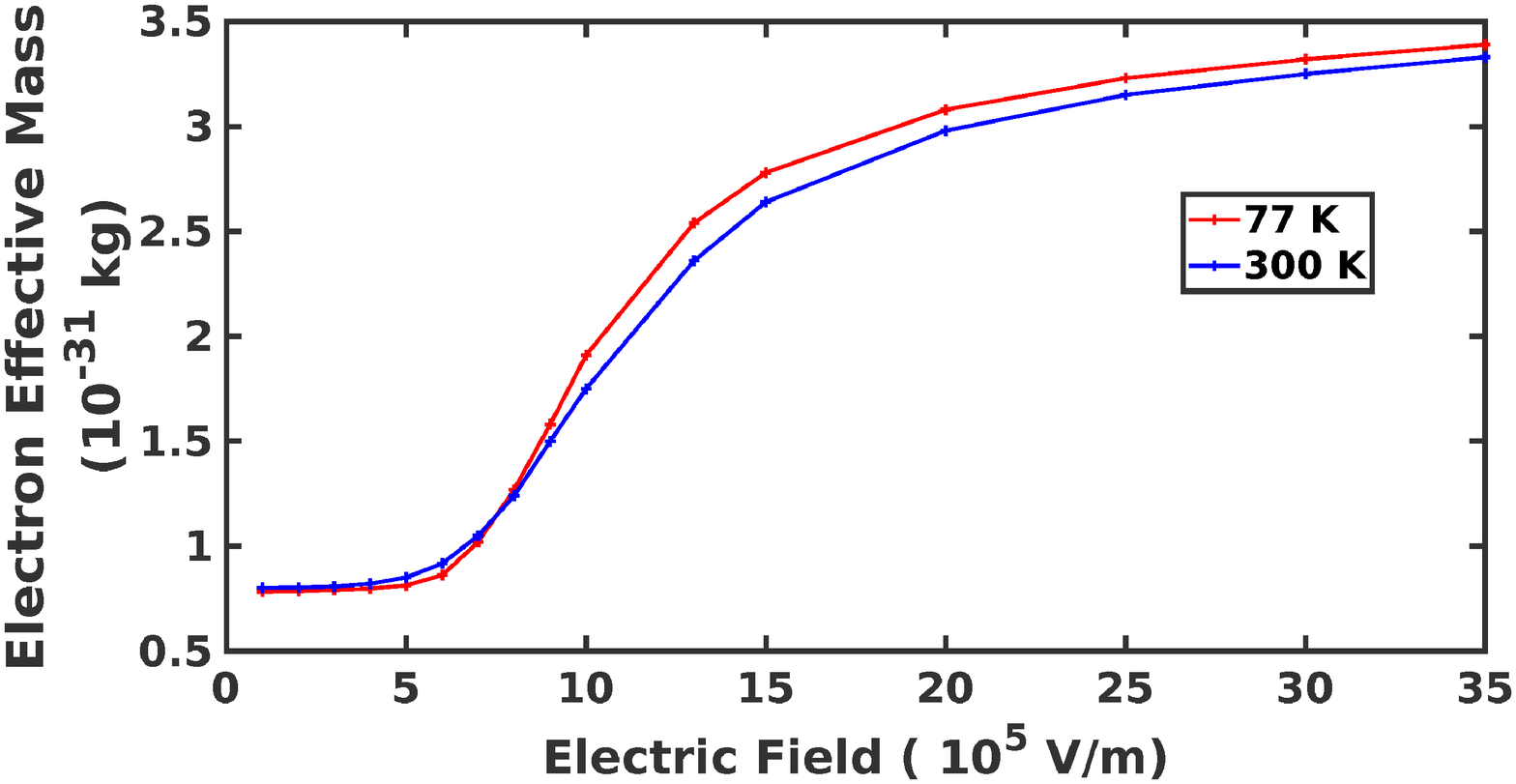} }
\subfigure[$ $]{\includegraphics[height=0.37\textwidth,width=0.48\textwidth]{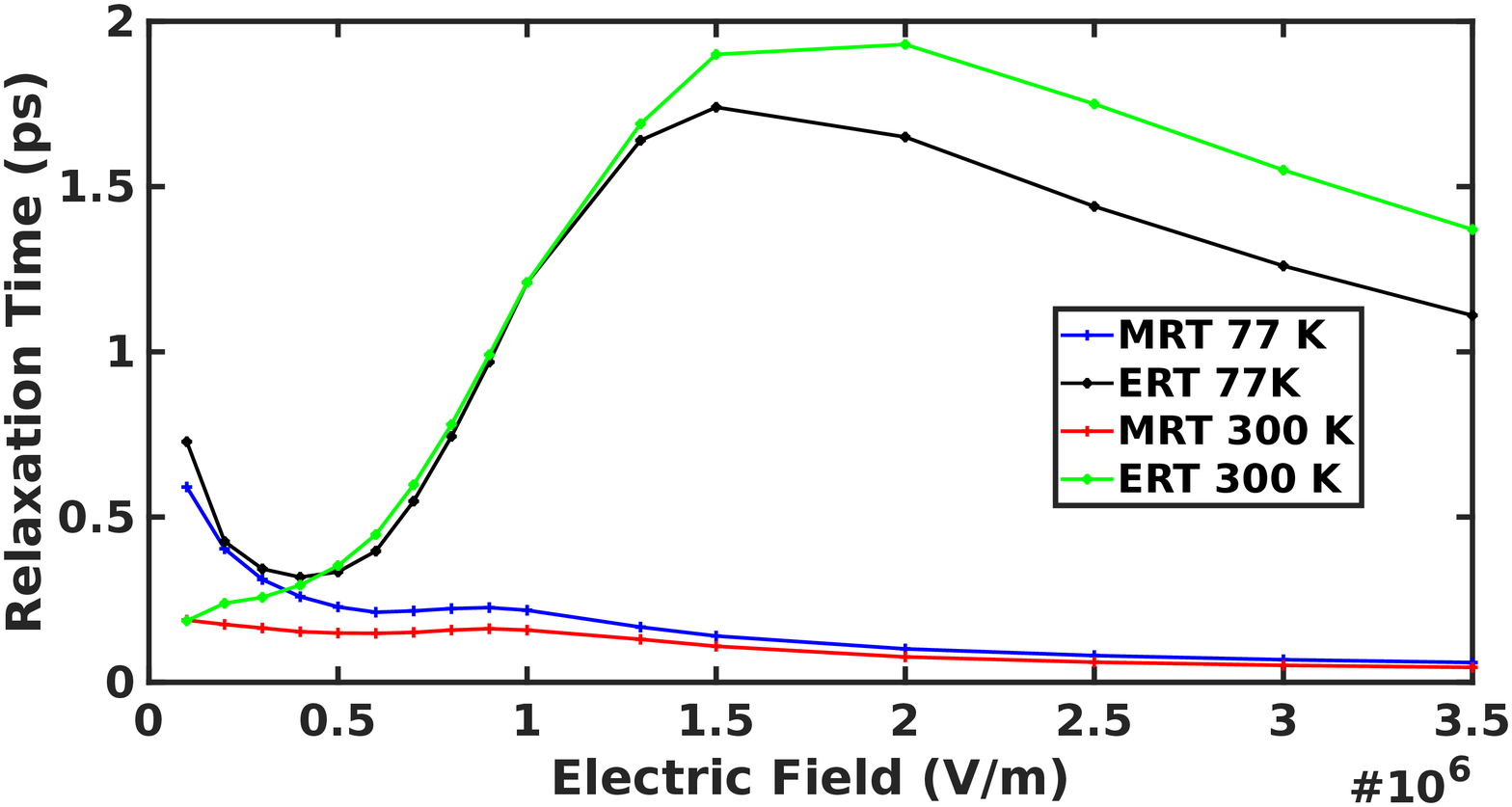} }
\caption{(a) Electron effective mass as a function of electric field at $77$ K and $300$ K (b) Momentum and energy relaxation time as a function of electric field at $77$ K and $300$ K}
\label{fig19}
\end{figure}

Figure \ref{fig19} (a) shows the variation of electron effective mass with electric field. With increasing electric field, average electron energy increases and electron are shifted to higher energy region in the same valley or to the satellite L valley from gamma valley. Both these factors lead to increase in electron effective mass, since electron effective mass is given by $m_{eff} = m^*(1 + 2 \alpha E)$, so with increasing electric field, electrons gets shifted to higher energy region in the same valley so its effective mass also increases and second if electrons are shifted to L valley then, L-valley also have higher effective mass than gamma valley. In low field region electron effective mass remains almost fixed with increasing electric field. In between region of $5 \times 10^5 $ V/m to $25 \times 10^5 $ V/m electrical field, there is a significant increase in electron effective mass with increasing electric field, since due to inter-valley scattering lots of electrons are shifted to L valley from the gamma valley in this region. At higher electric field after $25 \times 10^5 $ V/m, there is a very slow increase in electron effective mass, since as also depicted in Fig. \ref{fig2} (a) after $25 \times 10^5$ V/m there is a slight increase in L valley occupancy.    

Figure \ref{fig19} (b) shows the variation of momentum and energy relaxation time with electric field at $77$ K and $300$ K. Momentum relaxation time decreases with increasing electric field. At lower electric field most of the electrons are in the gamma valley and intervalley scattering will not play the main role. In the absence of intervalley scattering at lower electric field, electrons would relax its momentum over longer time and results in higher momentum relaxation time. At higher electric field intervalley scattering scattering become important and electron relaxes momentum at a faster rate due to higher effective mass and the higher scattering rate in the upper valley. \\

At $300$ K in the low field region in between $1 \times 10^5 $ V/m to $20 \times 10^5 $ V/m energy relaxation time increases with electric field and then start decreasing with electric field. In low field region in between $5 \times 10^5 $ V/m to $20 \times 10^5 $ V/m as show in Fig. \ref{fig2} (b) average electron energy increases with electric field significantly then it increases at a slower rate. In the lower field region, most of the electrons are in the gamma valley and energy is relaxed by mainly emitting polar optical phonon. However, little energy of electrons is relaxed by emitting polar optical phonon scattering in this region, results in increase of average electron energy and energy relaxation time sharply. At higher electric field intervalley scattering will begin to play an important role and second at higher electric field most of the electrons are in the higher L-valley, which has a much higher total scattering rate than the gamma valley total scattering rate, so energy relaxation time start decreasing at higher electric field. 


\subsubsection{Diffusion Coefficient}

\begin{figure}
\includegraphics[height=0.37\textwidth,width=0.48\textwidth]{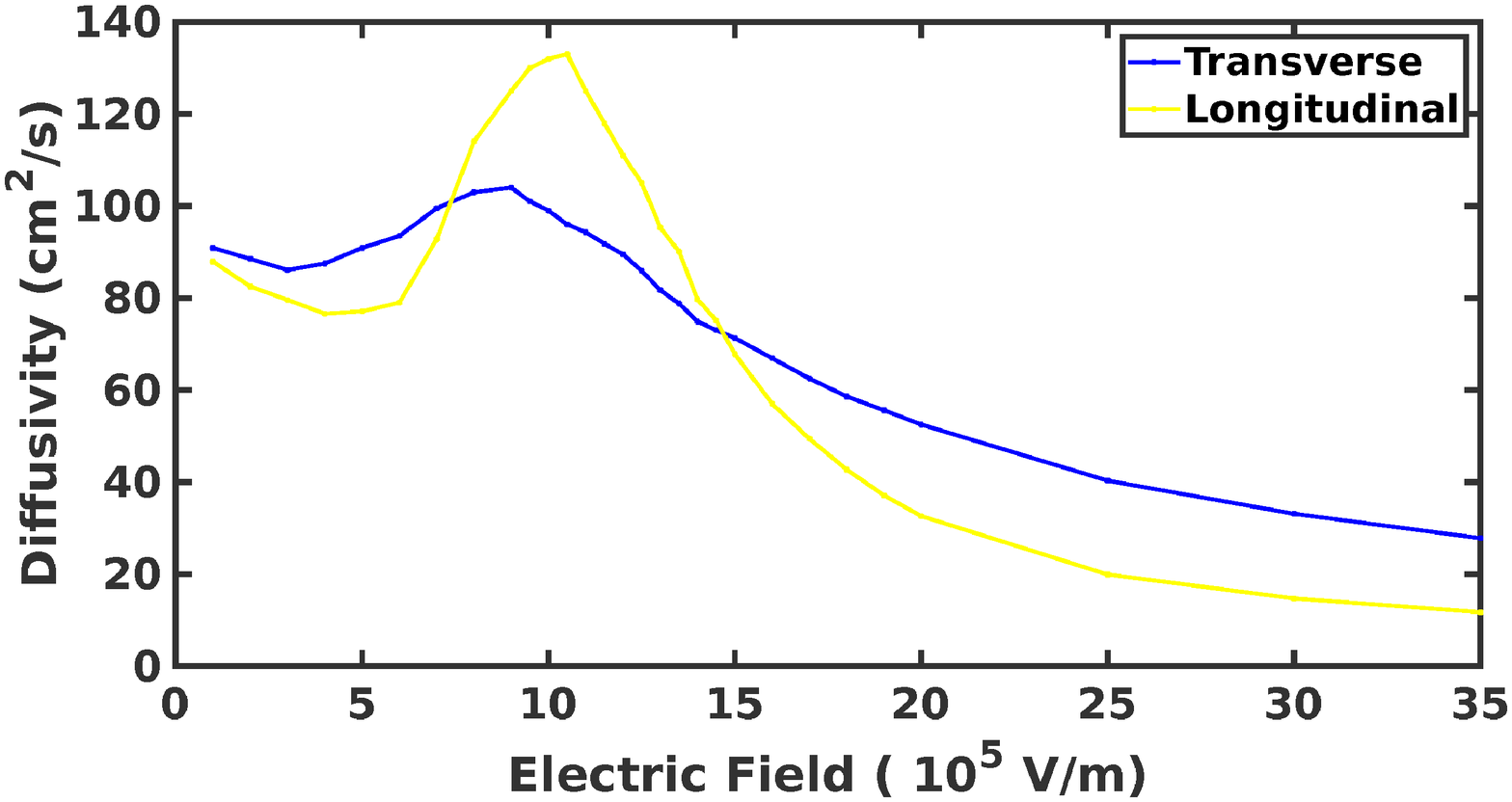} 
\caption{ Diffusivity as a function of electric field at $300$ K }
\label{fig18}
\end{figure}

Figure \ref{fig18} shows the variation of longitudinal and transverse diffusion coefficient with electric field at $300$ K. Both longitudinal and transverse diffusion coefficient shows a peak in diffusivity near critical electrical field. Below or around the critical electrical field due to rapid increase in electron energy raises the diffusion coefficient. At well higher electric fields above critical field, average electron energy increases slowly and drift velocity and mobility reduces since due to intervalley scattering electrons shifted to higher effective mass satellite valley. So, at higher electric field diffusion coefficient reduces to very low values. Saturated drift velocity and low diffusion coefficient at higher electric field may have remarkable implications for high frequency device operations. The smaller values of longitudinal and transverse diffusion coefficient at higher electric field leads to lower diffusion noise. So, lower diffusion noise can be achieved by applying the higher electric fields without loss of speed. The anisotropy between longitudinal and transfer, diffusion coefficient here is lower than observed in InP \cite{InP} and CdTe\cite{CdTe}, and of approx same magnitude as GaAs\cite{GaAs}. This reflects a lower energy separation between central and satellite valley of In\textsubscript{0.52}Al\textsubscript{0.48}As (0.34 eV) than InP (0.52 eV )and CdTe (1.5 eV ) and of approx same magnitude as of GaAs (0.35 eV) and it supports the explanation of the difference between the two coefficients given in Ref. \cite{CdTe} and \cite{GaAs}.


\subsubsection{Transient Electron Transport}

\begin{figure}
\subfigure[$ $]{\includegraphics[height=0.37\textwidth,width=0.48\textwidth]{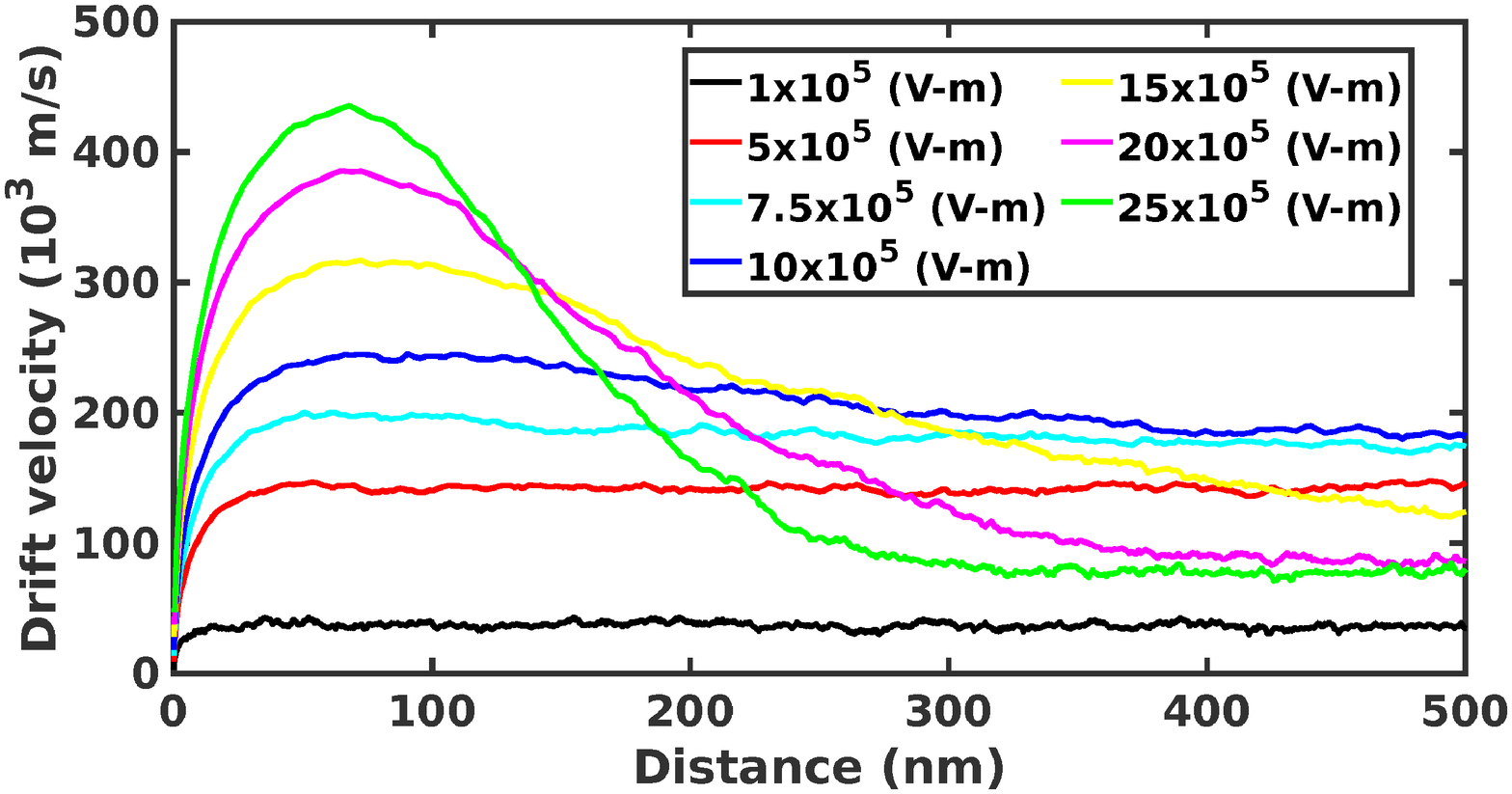} }
\subfigure[$ $]{\includegraphics[height=0.37\textwidth,width=0.48\textwidth]{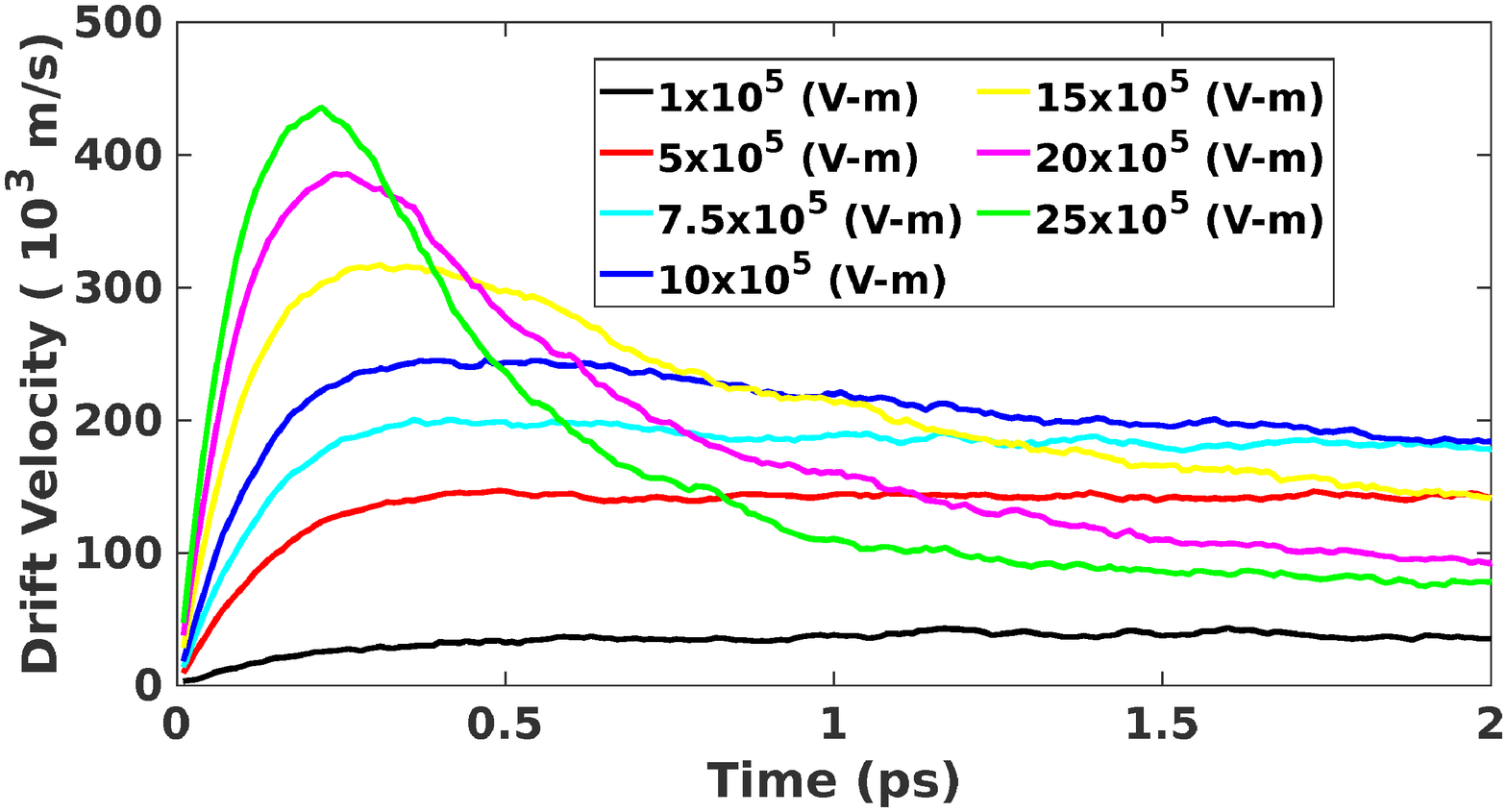} }
\caption{(a) Drift velocity as a function of distance displaced, for various applied electric field strength. (b) Drift velocity as a function of time elapsed since the application of electric field, for various applied electric field strength. For all cases temperature is set to $300$ K }
\label{fig11}
\end{figure}

We now examine the transient electron transport of In\textsubscript{0.52}Al\textsubscript{0.48}As. Figure \ref{fig11} (a) shows the electron drift velocity as a function of the distance traveled since the application of electric field for various applied electric field strength at $ 300 $ K. For applied field upto $7.5 \times 10^5 $ V/m electron reaches steady state very quickly with little or no velocity overshoot. For applied electric field higher than $7.5 \times 10^5 $ V/m significant velocity overshoot occurs. This result suggests that for In\textsubscript{0.52}Al\textsubscript{0.48}As $7.5 \times 10^5 $ V/m is critical applied field strength for the onset of velocity overshoot effects. At $300$ K it is already mentioned that $7.5 \times 10^5 $ V/m is corresponds to the electric field for peak drift velocity. Similar results found for GaN, ZnO and other III-V semiconductors \cite{ZnO,GaN}.

\begin{figure}
\subfigure[$ $]{\includegraphics[height=0.37\textwidth,width=0.48\textwidth]{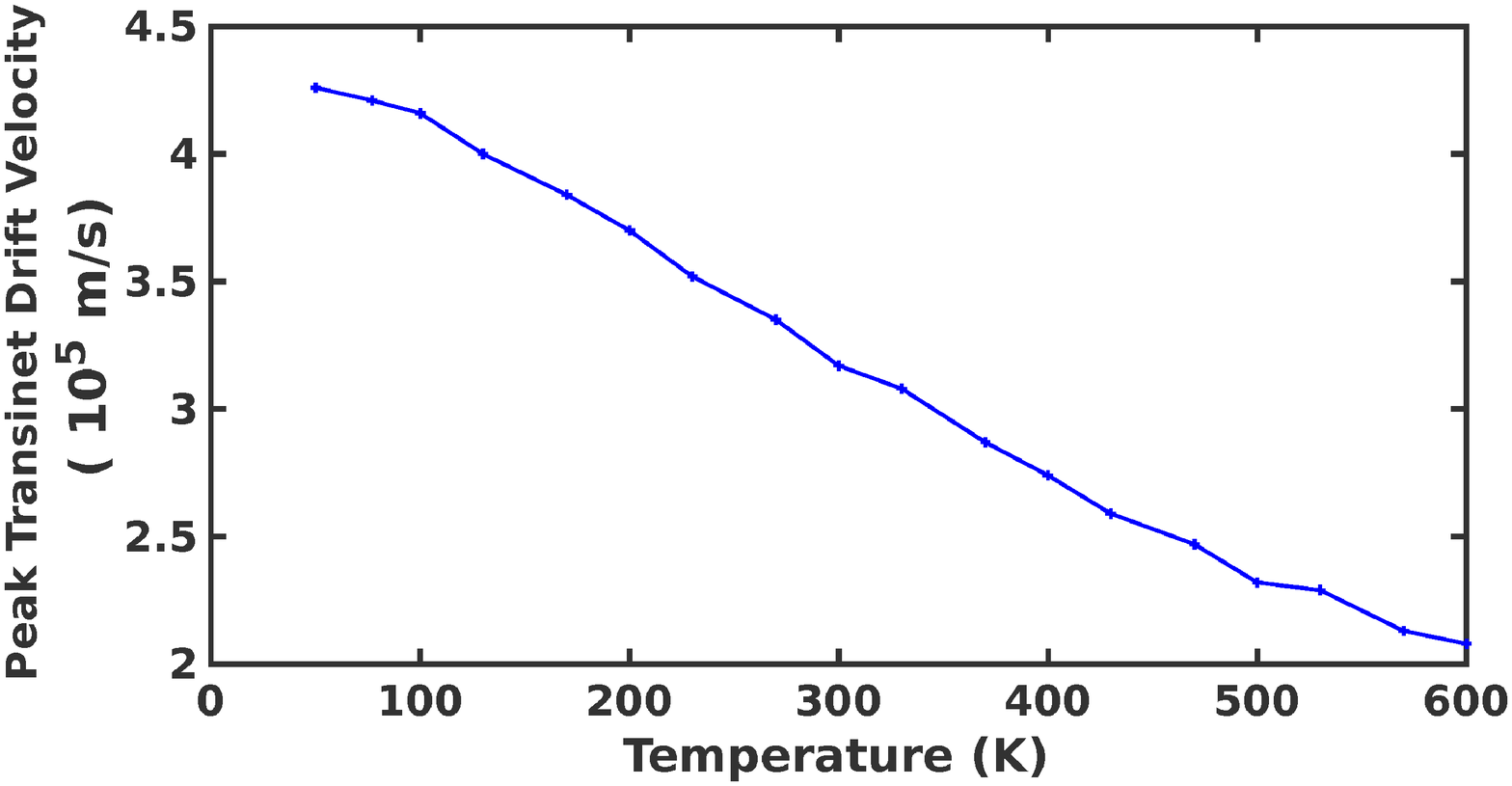} }
\subfigure[$ $]{\includegraphics[height=0.37\textwidth,width=0.48\textwidth]{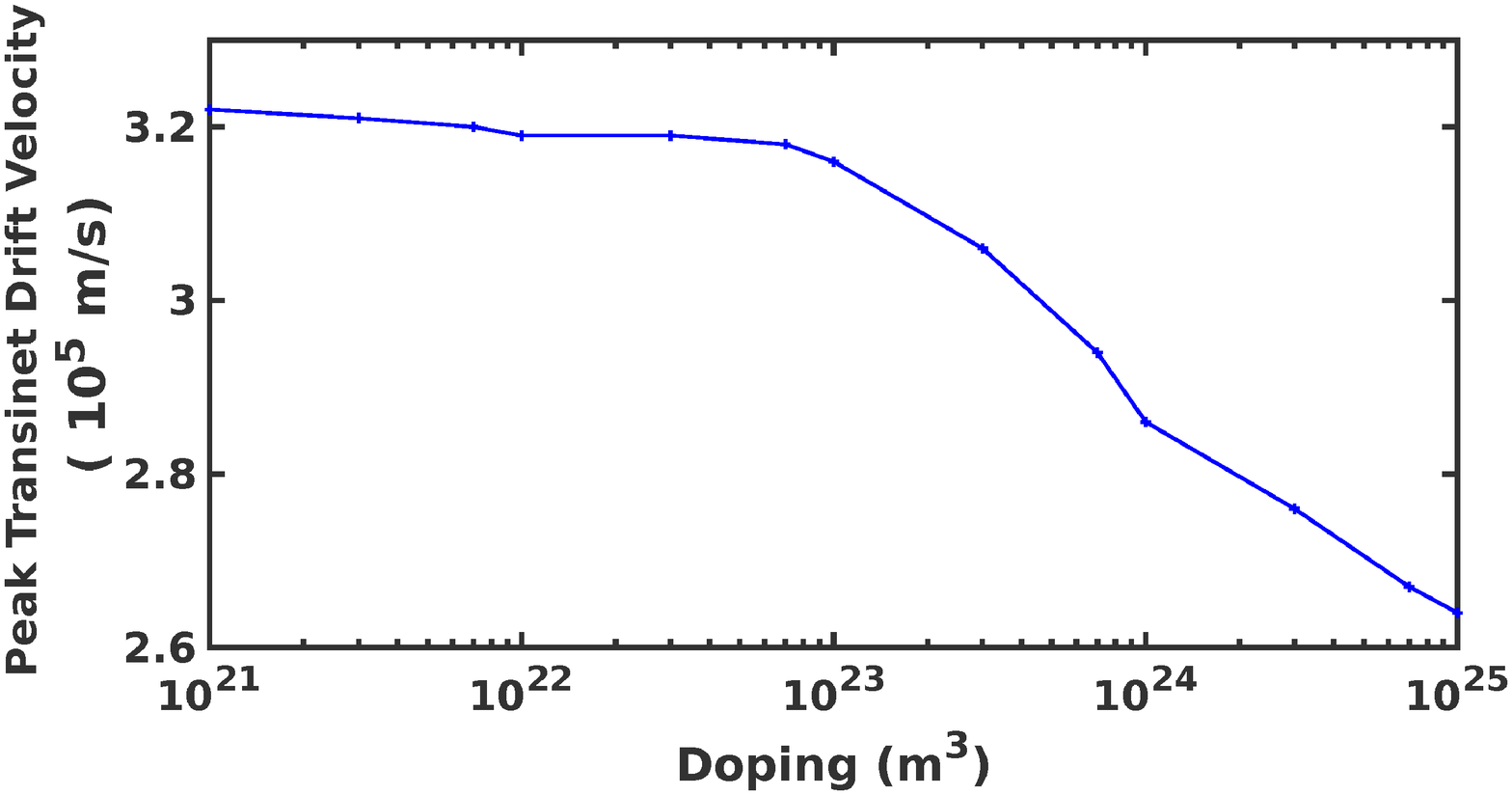} }
\caption{(a) Peak transient drift velocity as a function of temperature (b) Peak transient drift velocity as a function of doping concentration}
\label{peak_transient}
\end{figure}

Figure \ref{fig11} (b) shows the drift velocity variation with time. It has also same trend as the Fig. \ref{fig11} (a). For upto $7.5 \times 10^5 $ V/m electric field, there is very little or no overshoot. For electric field higher that $7.5 \times 10^5 $ V/m there is significant velocity overshoot occur. Figure \ref{peak_transient} (a) shows the variation of peak transient drift velocity as function of temperature and Fig. \ref{peak_transient} (b) shows the variation of peak transient drift velocity with doping concentration, for both cases the applied electric field strength being set to $15 \times 10^5 $ V/m.

\begin{figure}
\includegraphics[height=0.37\textwidth,width=0.48\textwidth]{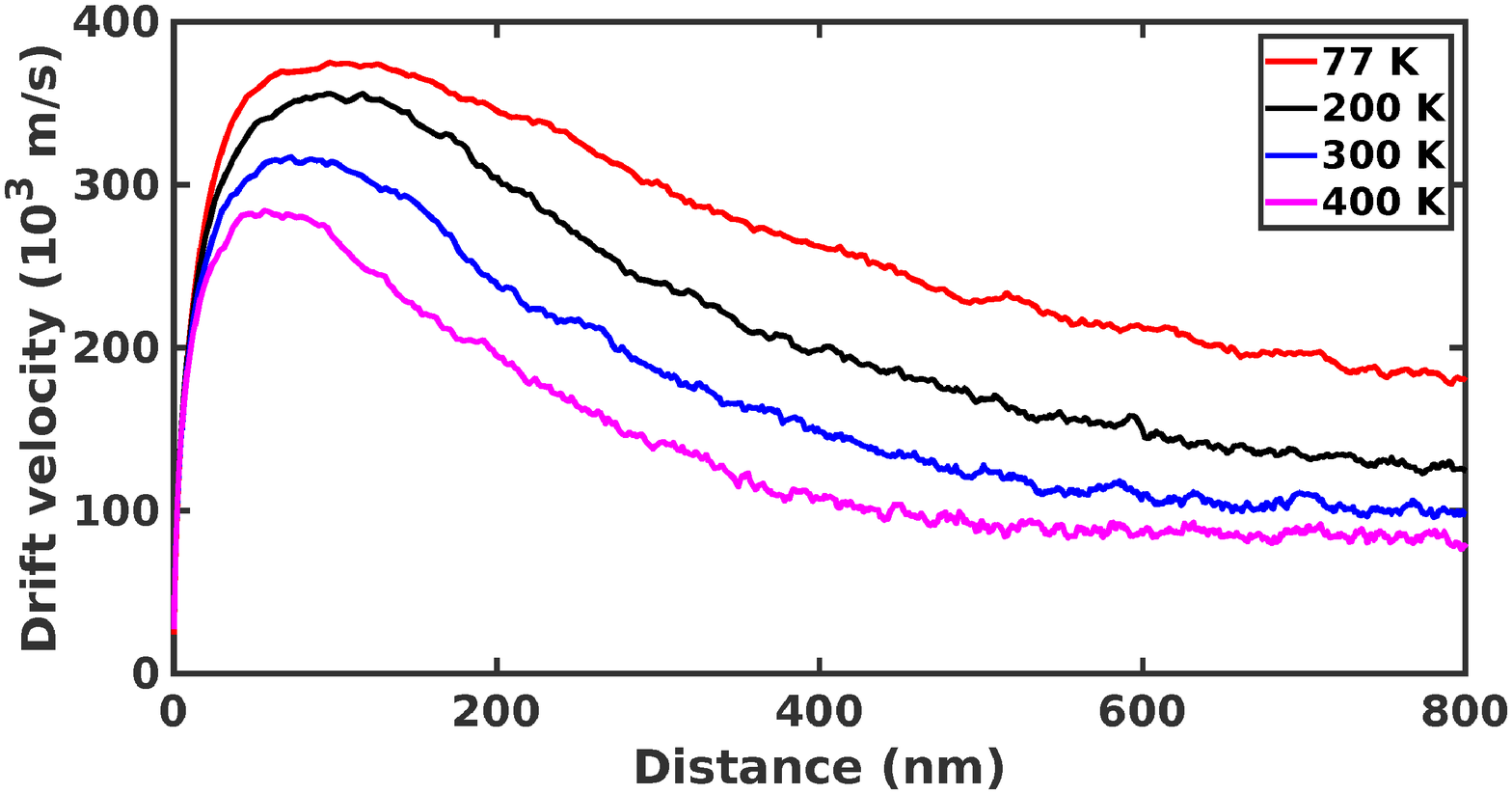} 
\caption{ Electron drift velocity as a function of distance displaced since the application of electric field at different temperatures. For all cases electric field is set to two times the critical electric field}
\label{fig13}
\end{figure}

Figure \ref{fig13} shows the variation of drift velocity with distance for different temperatures. We have followed the same approach as in the paper \cite{ZnO} and set the electric field twice the approximate critical field for each case. The critical field for temperatures $77$ K, $200$ K, $300$ K and $400$ K are $6 \times 10^5 $ V/m, $7 \times 10^5 $ V/m, $7.5 \times 10^5 $ V/m and $8 \times 10^5 $ V/m respectively. Crystal temperature has significant effect on transient electron transport. Peak drift velocity is about $371$ m/s when temperature is $77$ K and it reduces to about $281$ m/s when crystal temperature is about $ 400$ K. For higher crystal temperatures steady state is achieved at much higher rate.

\begin{figure}
\includegraphics[height=0.37\textwidth,width=0.48\textwidth]{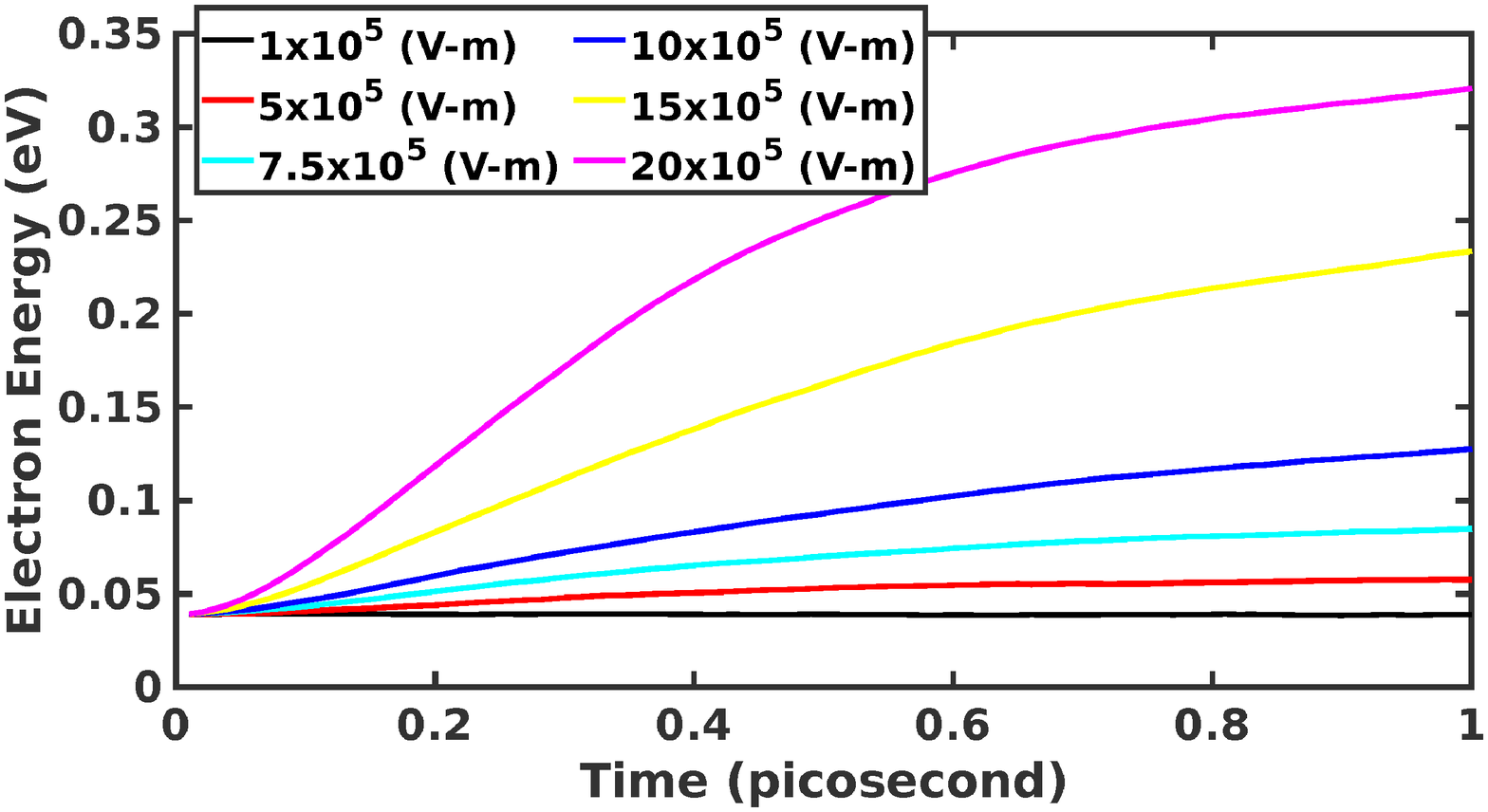} 
\caption{ Electron energy as a function of time elapsed since the application of electric field, for various applied electric field strength. For all cases temperature is set to $300$ K  }
\label{fig14}
\end{figure}

Figure \ref{fig14} shows the variation of electron energy with time for different applied electric field at $300$ K. Electron energy increases monotonically with time for all applied electric fields till it reaches steady state. For low applied electric field steady state reaches very quickly.

\begin{figure}
\includegraphics[height=0.37\textwidth,width=0.48\textwidth]{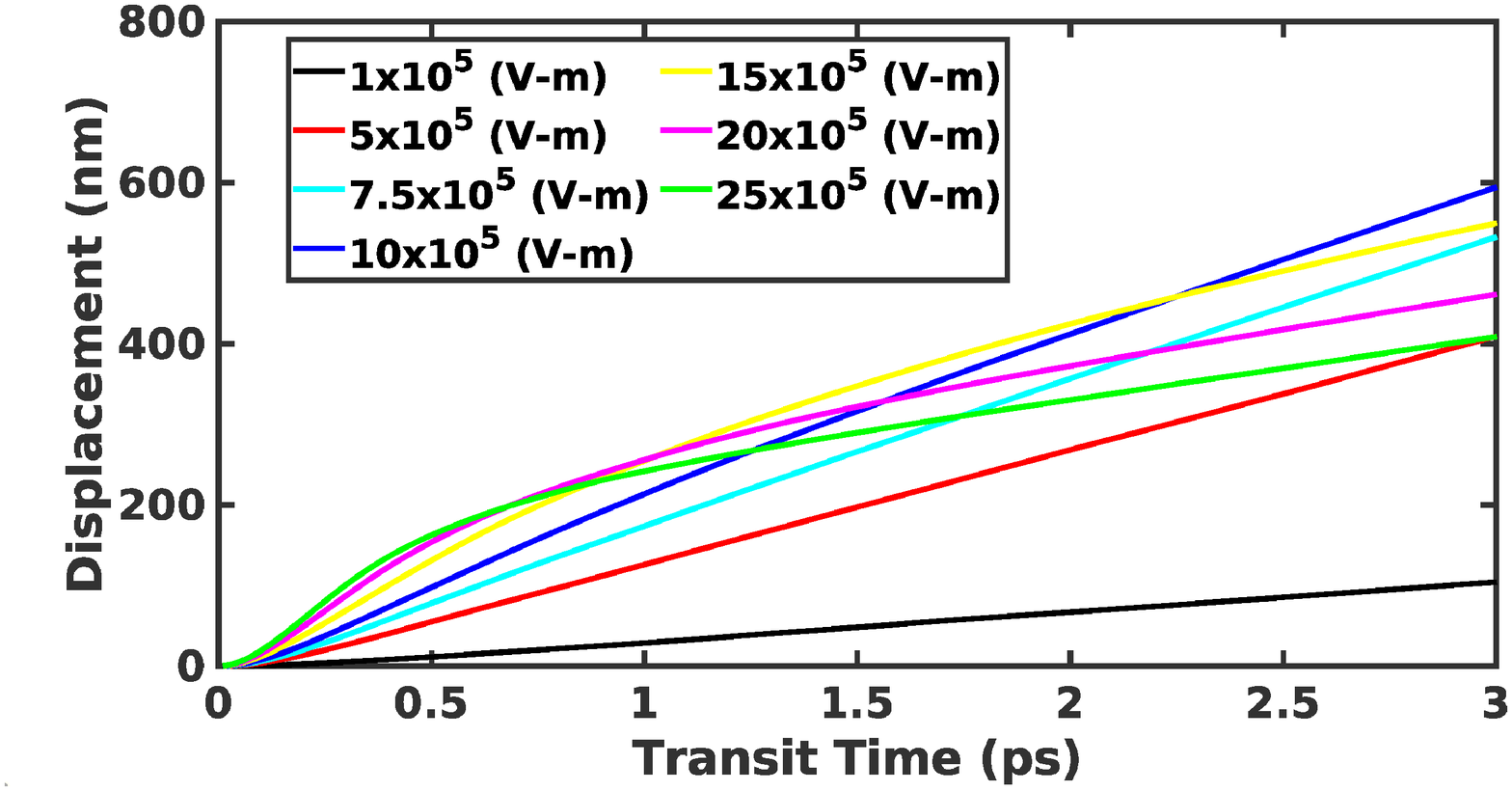} 
\caption{Electron displacement as a function of transit time for various applied field strength at $300$ K }
\label{fig15}
\end{figure}

Figure \ref{fig15} shows the variation of electron displacement as a function of time elapsed since the application of electric field for a number of different cases. Electron displacement increases monotonically in response to increase in time elapsed since the onset of applied electric field.     

\subsection{Device Implications}
\begin{figure}
\includegraphics[height=0.37\textwidth,width=0.48\textwidth]{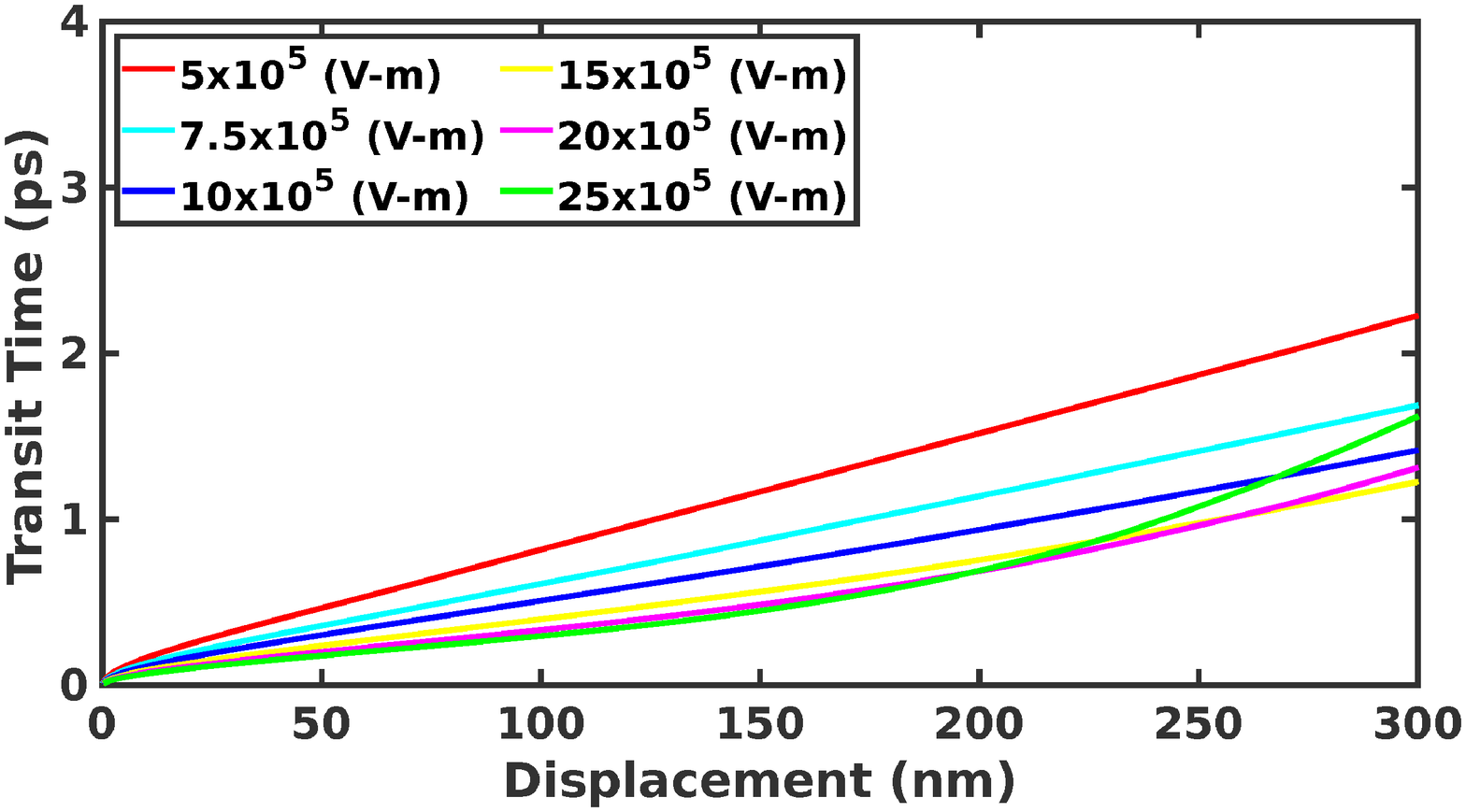} 
\caption{ Electron transit time as a function of distance displaced for various applied electric field strength at $300$ K }
\label{fig21}
\end{figure}

\begin{figure}
\includegraphics[height=0.37\textwidth,width=0.48\textwidth]{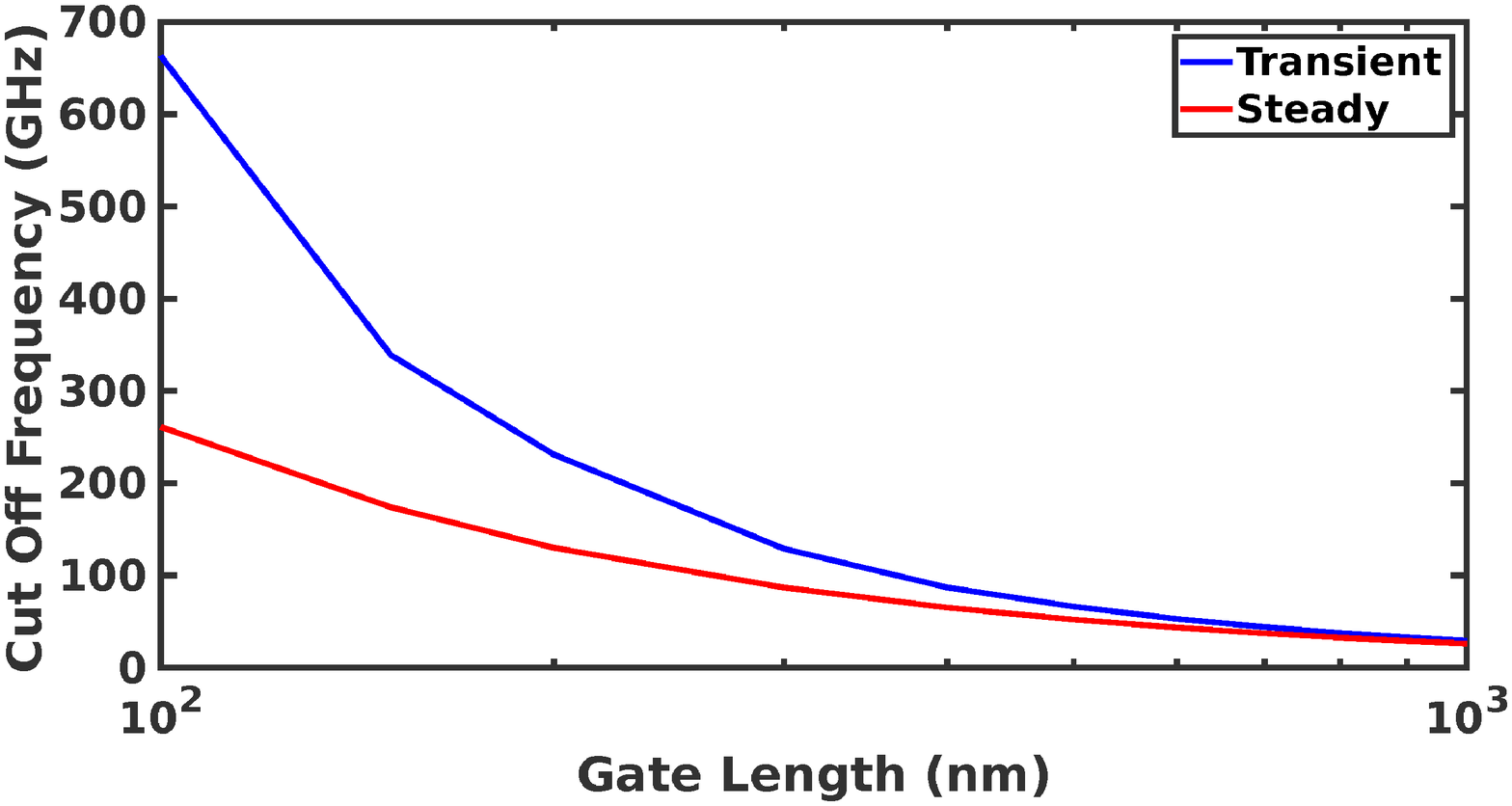} 
\caption{ The optimal cut-off frequency as a function of device gate length. Blue colour curve is obtained by including velocity overshoot effect and red colour curve is obtained without including velocity overshoot effect }
\label{fig22}
\end{figure}

The transient electron transport that we have studied till now can be used to enhance the performance of electron device fabricated from In\textsubscript{0.52}Al\textsubscript{0.48}As. Note that the upper bound on the cut-off frequency of a device is given by the formula 

\begin{equation}
f_T = \frac{1}{2\pi \tau}
\label{ft}
\end{equation}

where $\tau$ is transit time across the device. To determine an upper bound for cut-off frequency, first we have to determine the minimum transit time occurring for optimally chosen applied field. In Fig. \ref{fig21} we have plotted the average transit time as a function of displacement for different applied electric field. In this curve for a displacement of $400$ nm, minimum transit time is obtained with an electric field of $15 \times 10^5 $ V/m. Similarly, we have calculated the minimum transit time required for the different device length of In\textsubscript{0.52}Al\textsubscript{0.48}As material. From this optimization procedure, we have calculated the upper bound on the cut-off frequency for different device length and plotted in Fig. \ref{fig22}. The blue color curve in Fig. \ref{fig22} represents the optimize results obtained by incorporating the velocity overshot effect occurring during the transient state of electron transport. While the red curve in Fig. \ref{fig22} represents the upper bound on cut-off frequency obtained by applying the field which produces largest steady state electron velocity, i.e. this curve does not include the effect of transient state. At lower device length there is significant improvement in upper bound on the cut-off frequency can be obtained due to velocity overshoot effect. For device length smaller than $ 700$ nm, transient effect becomes noticeable and it becomes more pronounced as device length is diminished further. For device length of $100$ nm upper bound on cut-off frequency is improved from $261$ GHz to $663 $ GHz by including transient effect into calculation. While doing the calculation of upper bound on cut-off frequency all non-idealities occurring during normal device operation are ignored.

\section{Conclusion}

We presented a detailed and comprehensive study of steady state and transient electronic transport in In\textsubscript{0.52}Al\textsubscript{0.48}As with the three valley model using the semi-classical ensemble Monte Carlo method and including all important scattering mechanisms. All electronic transport parameters such drift velocity, valley occupation, average electron energy, ionization coefficient and generation rate, electron effective mass, diffusion coefficient, energy and momentum relaxation time were extracted rigorously from the simulations. Using these, we presented a complete characterization of the transient electronic transport showing the variation of drift velocity with distance and time. If the applied electric field is higher than threshold field $ 7.5 \times 10^5 $ V/m for peak drift velocity, then velocity overshoot is observed during transient state. Transient effects becomes more pronounced at shorter device length. We then estimated the optimal cut-off frequencies for various device lengths via the velocity overshoot effect. Our analysis showed that for device lengths shorter than $700$ nm, transient effects are significant and should be taken into account for optimal device designs. As a critical example, at length scales of around $100$ nm, we obtained a significant improvement in the cut-off frequency from $261$ GHz to $663$ GHz with the inclusion of transient effects. The field dependence of all extracted parameters here could prove to be helpful for further device analysis and design.

{\it{Acknowledgements:}} AKM and BM gratefully acknowledge funding from Indo-Korea Science and Technology Center (IKST), Bangalore.


\bibliographystyle{apsrev}
\bibliography{Reference.bib}

\end{document}